\documentclass[number]{elsarticle}
\usepackage[utf8]{inputenc}
\usepackage{amsmath}
\usepackage{amsthm}
\usepackage{graphicx}
\usepackage{booktabs}
\usepackage{subfig}
\usepackage{amssymb}
\usepackage{algorithm}
\usepackage{algpseudocode}
\usepackage{hyperref}
\usepackage{xcolor}

\usepackage{url}
\newtheorem{theorem}{Theorem}
\newtheorem{lemma}[theorem]{Lemma}

\newproof{prf}{Proof}

\begin{document}

\title{Budget-balanced and strategy-proof auctions for ridesharing}

\author[1]{Leonardo Y. Schwarzstein}
\ead{leoyvens@gmail.com}
\author[2]{Rafael C. S. Schouery\corref{cor1}}
\ead{rafael@ic.unicamp.br}

\address{Institute of Computing, University of Campinas, Av. Albert Einstein 1251, Campinas 13083--852, Brazil}

\cortext[cor1]{Corresponding author}

\begin{abstract}
    Ridesharing services have become widespread, and pricing the rides is a crucial problem for these systems. We propose and analyze a budget-balanced and strategy-proof auction, the Weighted Minimum Surplus (WMS) auction, for the dynamic ridesharing problem with multiple passengers per ride. Under the assumption of downward closed alternatives, we obtain lower bounds for the surplus welfare and surplus profit of the WMS auction. We also propose and analyze a budget-balanced version of the well-known VCG mechanism, the~$\mathrm{VCG}_s$. Encouraging experimental results were obtained for both the WMS auction and the $\mathrm{VCG}_s$.
\end{abstract}


\maketitle

\section{Introduction}

The prevalence of smartphones equipped with internet connectivity and GPS systems has enabled the rapid growth of platforms that allow passengers to make a short-term request for the service of independent drivers. When the driver's motivation is to obtain a recurring income from these trips, such as in the platforms Uber and Lyft, this is called \emph{ridesourcing}~\cite{rayle2014app}, which has its own challenges when considering how to price users~\cite{SunRZG19}. This scenario can be distinguished from \emph{ridesharing}, where the driver seeks to share their trip costs by serving passengers that have a route similar to their own. Waze Carpool~\cite{Waze} is an example of a commercial platform that facilitates ridesharing. Some ridesourcing platforms also seek sharing as a way to reduce costs, for example, UberPool~\cite{Uber} and Lyft Line offer a smaller price by allowing the trip to be shared among multiple passengers. The InDriver~\cite{InDriver} platform resembles an auction since it allows passengers to advertise rides and prices, to which drivers can then give a counteroffer. 

Beyond the benefits for the participants of the trip, ridesharing is advantageous for the society as a whole, as it may reduce the congestion and pollutant emission caused by the use of personal cars in metropolitan areas by increasing the occupancy rate of vehicles and reducing the need for parking~\cite{hahn2017ridesharing,Guo23}. As an example, the metropolitan area of São Paulo, one of the largest in the world, is in a long-lasting mobility crisis that was estimated to cost up to 1\% of Brazilian GDP in 2012~\cite{cintra2014custos}, with congestion resulting in up to three times more emissions of pollutants than the expected emission without congestion~\cite{cintra2014custos}.
These systems can also be combined with public transportation systems in order to provide reliable and affordable transit, while reducing CO2 emissions, especially for residents in suburban areas~\cite{Mitja18,Yu21,Asghari22}.

In this paper, we are particularly interested in \emph{dynamic ridesharing} with trips that may be shared among multiple passengers, which allows drivers to serve more passengers and incurs lower costs for the passengers and the society. The \emph{Dynamic Ridesharing Problem}, or real-time ridesharing, is characterized by the following properties: \emph{Dynamic}, the trips must be formed rapidly as new requests come in continuously; \emph{Independent}, drivers and passengers are independent agents and therefore will only share a trip if that is beneficial to them; \emph{Cost sharing}, participants wish to obtain a lower cost than the one incurred by traveling alone; \emph{Non-recurring trips}, single short-term trips rather than recurring pre-arranged trips; \emph{Automatic}, the system should require minimal operational effort from the participants~\cite{agatz_optimization_2012}.
Note that, in this paper, the term \emph{dynamic} is used to define a ridesharing scenario with the above properties instead of denoting that the driver's route can change as other passengers requests the service as in the works of Sayarshad et al.~\cite{sayarshad2015scalable} and Bian et al.~\cite{BIAN202077}.

Dynamic ridesharing involves complex routing and pricing aspects. The \emph{Dial-A-Ride Problem}~(DARP)~\cite{cordeau_branch-and-cut_2003} is a good routing model for multi-passenger ridesharing, where passengers have pick-up and drop-off locations that may not coincide with the driver's initial location and destination, while also having constraints on pick up time and travel time. For a review on Dial-A-Ride problems, please confer the article from Molenbruch et al.~\cite{Molenbruch2017}.

Cost minimization routing problems have received much attention in the operations research literature~\cite{pillac2013review}, for example, Santos and Xavier~\cite{santos2013dynamic} present heuristics to maximize taxi sharing in a DARP problem. The ridesharing problem with dynamic pricing brings new challenges to routing problems, such as balancing the number of passengers served and the profit obtained. For example, even if we have only one driver and two passengers, there are instances where any way to share the costs between the driver and passengers will not be in the core, an important concept of fairness in cooperative game theory.

Trip pricing gains importance in dynamic ridesharing because the independence of the agents makes it necessary to incentivize the participation of passengers and, especially, drivers in the system. However, pricing has received less attention from the ridesharing literature~\cite{furuhata_ridesharing:_2013}. In moments of high demand, when there are few drivers and many passengers, it is necessary to raise prices to attract drivers into the system and balance supply and demand. In this scenario, cost minimization is no longer the main goal of the system's optimization, since the willingness to pay (or budget) of each passenger for the trip must be considered. Some ridesourcing platforms such as Uber and Lyft raise the price per distance in these scenarios, a policy known as \emph{surge pricing}~\cite{GudaS19,Garg22}.
Current studies suggest that surge pricing achieves its goals of improving economic efficiency~\cite{hahn2017ridesharing}. Still, surge pricing has been viewed negatively by consumers and regulators~\cite{cachon_surge}. When the system decides on a higher price, consumers might perceive that it is acting against them. In an auction, however, the consumers' bids would determine the price, so we explore auctions as a form of balancing the market in moments of high demand with a price that is better justifiable to the consumer.

\subsection{Literature Review}

Two desirable properties of auctions that are studied in previous literature on auction-based pricing for ridesharing are that of (weak) \emph{budget-balance}, which means the price paid by users is at least the driver's cost for serving them, and \emph{strategy-proofness}, which means the participants do not profit by misreporting information to the auction. Kamar and Horvitz~\cite{kamar_collaboration_2009} propose a multi-driver system based on a Vickrey-Clark-Groves (VCG)~\cite{vickrey_counterspeculation_1961, clarke1971multipart, groves_incentives_1973} auction. The VCG auction is not budget-balanced and, due to the use of heuristics for better computational performance, is implemented in a way that also is not strategy-proof. Kleiner et al.~\cite{kleiner_mechanism_2011} uses a strategy-proof auction adapted from the second-price auction~\cite{vickrey_counterspeculation_1961}, where each driver may serve a single passenger. Considering each driver in isolation, this auction is strategy-proof and budget-balanced, as long as the driver is not allowed to choose the passenger to be served. Our work extends this line of research by proposing auctions that are multi-passenger, strategy-proof, and budget-balanced.

Another line of research involves making prior assumptions on the passengers' arrival rate and value distributions, and then calculating prices that are optimal in expectation. Sayarshad and Chow~\cite{sayarshad2015scalable} and Chen et al.~\cite{chen_optimal} explore this type of mechanism, while Masoud et al.~\cite{masoud2017using} assumes a prior distribution on the passengers' values to propose an optimal price for which a passenger may buy a trip from another passenger.

In recent literature, Zhang et al.~\cite{zhang2016discounted} apply a bilateral trade reduction mechanism that is strategy-proof, modifying the McAfee mechanism~\cite{mcafee1992dominant} for a ridesourcing scenario. A limitation of this mechanism is that a driver may only serve one passenger at a time, and budget balance is not guaranteed. Zhao~et~al.~\cite{ZhaoVCG} propose a multi-driver and multi-passenger theoretical model that uses a VCG auction with reserve prices, but they achieve budget balance only when the drivers make no detours, while the auctions we propose are budget-balanced even with detours. Shen et al.~\cite{shen2016online} propose an online ridesharing system where a passenger that enters the system is offered a price estimate to be served by an available driver, assuming a scenario where the driver supply is sufficient for passengers to be immediately served. The authors claim that the system is strategy-proof, however, the final price may be lower than the accepted estimate, so the passenger may strategically accept a higher estimate hoping to get a lower final price. This system is in fact individually rational, that is, it guarantees that a passenger that only accepts estimates lower or equal to their value will not suffer a loss.
Bian et al.~\cite{BIAN202077} consider a rolling horizon planning approach to serve users of a first-mile ridesharing system, accounting for passengers' mobility preferences such as deadlines and detour tolerance. Their model allows for changes in the drivers' route to pick up new passengers that arrive later on in the system. For this, they propose a VCG-based mechanism that is individually rational, incentive compatible, and has price controllability (guaranteeing that at least a baseline price will be paid). Finally, they present a heuristic for solving large-scale ridesharing auctions.

Ma et al.~\cite{MaFP20} presented the Spatio-Temporal Pricing mechanism, which considers a multi-driver system for platforms such as Uber and Lyft. The mechanism incentivizes the drivers to accept the passengers' requests independently of location or time, ensures that the driver do not envy the payment made to other drivers and passengers do not envy the price paid by other passengers, among other interesting guarantees, while maximizing the social fare (in this case, the passengers values minus the drivers costs). We note that their work differs from ours as they consider a single passenger per trip, while we consider that more than one passenger can be at the driver's vehicle at the same time.

Fieldbaum et at.~\cite{Fielbaum22} models ridesharing as a cooperative game, and propose a cost-sharing protocol that makes the optimal solution an equilibrium for the participants. This is a different approach than ours, as there are no bids, and one must only share the costs of transportation between the users.

A comprehensive survey by Furuhata et al.~\cite{furuhata_ridesharing:_2013} classifies ridesharing as either for a single passenger or multiple passengers. It also describes four spatial patterns of increasing generality, of which the most general is detour ridesharing, where the driver may take detours and passenger pickup locations do not have to coincide with the driver's start location and neither do passenger delivery locations have to coincide with the driver's destination. This survey also identifies challenges for auction-based ridesharing, among them to find an auction that is strategy-proof, budget-balanced, and allows multiple passengers to be served. Even in recent literature, this challenge still has not been addressed.

\subsection{Our contribution}

In this paper, we propose auctions for single-driver, multi-passenger, detour ridesharing that are strategy-proof and (weakly) budget-balanced. 

In the considered scenario, a driver wants to reduce their travel cost by serving one or more passengers. For this, the passengers submit a bid, as well other relevant pieces of information such as pick-up and drop-off locations, to a system that selects the set of passengers to be served, decides how much each served passenger must pay, and defines the route of the driver. The bids are submitted before the driver leaves their starting location, and the costs of the driver for every possible route are public knowledge. The scenario is static, that is, it does not allow passengers to be added or removed from the set of served passengers during the driver's trip.

The main proposed auction is the \emph{Weighted Minimum Surplus}~(WMS) auction, for which we obtain a lower bound of the maximum social welfare and profit under certain conditions.
We also propose a budget-balanced version of the VCG auction, the $\mathrm{VCG}_s$ auction, which remains strategy-proof. We note that $\mathrm{VCG}_s$ have some similarities to the mechanism MPMBPC presented by Bian et al.~\cite{BIAN202077}. Nonetheless, $\mathrm{VCG}_s$ was developed independently, and they have some differences. Besides focusing on different problems, the price paid by the passengers is different and$\mathrm{VCG}_s$ is budget-balanced (that is, the driver cost is always paid in full) where MPMBPC has price controllability (passengers pays at least a baseline price, which is not, necessarily, sufficient to pay the drivers'~cost).

We present an experimental analysis of all proposed auctions, utilizing ridesharing instances generated on real-world maps. Both the WMS and the $\mathrm{VCG}_s$ auctions obtain high social welfare and achieve a good profit, particularly in instances with a larger number of passengers.

Though these auctions were analyzed only in single-driver scenarios, they may serve as a basis for generalization to multi-driver scenarios, which is a harder problem. They can be used for a multi-driver scenario, but will lose the strategy-proofness property due to the online nature of the multi-driver~problem.
\paragraph{Paper organization} This paper is organized as follows. In Section~\ref{sec:problem}, we present the proposed auction-based ridesharing system and, in Section~\ref{sec:auctions}, we present four different auctions that can be used in this system. Section~\ref{sec:bounds} presents some worst-case guarantees for our main auction.
In Section~\ref{sec:experiments}, we present our computational experiments. Finally, in Section~\ref{sec:multidriver}, we discuss the multiple driver scenario, presenting possible extensions of our auctions and the difficulties that arise in this case and, in Section~\ref{sec:conclusion}, we present our conclusions and possible future research topics.
\section{Auction-based dynamic ridesharing}\label{sec:problem}

In the proposed system, a driver with a specific destination wishes to share their trip with passengers in exchange for monetary gain or cost reduction. Each passenger wishes to be served, that is, transported from their pick-up location to their drop-off location.

Thus, our system is as follows. The driver posts their trip in the system, composed of information such as start and end locations and time constraints. Also, passengers submit their information (pick-up and drop-off locations, and time constraints, that is, any constraints that the system allows the passengers to set) along with a monetary value (a bid) for being served. The system collects these bids until a time limit (based on the driver's time constraints) and runs an auction to decide which passengers to serve and the price to be charged to every served passenger. The system then informs the result to the driver and the passengers, and the driver proceeds with their trip, picking up and delivering the selected passengers according to the route established by the~system.

The properties of the system depend on the auction used. Since there are multiple desirable properties for an auction studied in Auction Theory, we focus on four properties to evaluate auctions for this problem: being \emph{strategy-proof}, \emph{maximizing social welfare}, being (weakly) \emph{budget-balanced} and \emph{maximizing profit}.

Before introducing these properties, we must first define a trip. A \emph{trip} is a route that begins at the driver's start location, serves a set of passengers, and finishes at the driver's destination, respecting existing time and distance constraints.

In this paper, our experiments consider that passengers have a maximum pick-up time and a maximum travel time, and the driver has a maximum arrival time at the destination. We note that other constraints, such as distance constraints where the detour made by the driver or the passengers is limited, could be considered in the definition of a trip and in the respective optimization problem to be solved. Thus, our system can be used in different ridesharing scenarios.

Formally, a trip $A$ is represented simply by a set of served passengers, that is, $i \in A$ if and only if $i$ is served by trip $A$. The value~$\mathrm{\emph{cost}}(A)$ is the cost for the driver to serve~$A$, representing costs such as fuel and car wear and tear, and is public knowledge. The cost can, for example, be the driver's detour cost. The system must determine the winning trip $A^*$ to be served by the driver and the price $p_i$ to be charged from each passenger. Unserved passengers have~$p_i = 0$. Let $I$ be the set of all passengers, then the total value collected will be~$\sum_{i \in I} p_i$. Note that we call both served and non-served system users (except the driver) simply as passengers, even though some of them will not be picked up by the~driver.

We consider that every passenger $i$ has a value $v_i$, which represents the maximum value passenger $i$ is willing to pay to the system or, from another perspective, the monetary value that passenger $i$ gives for being served. We also consider that $v_i$ is private information, only known to passenger $i$.
The objective of a passenger in the auction is to maximize their \emph{utility} (or gain). The utility of passenger~$i \notin A^*$ is $u_i = 0$ since~${p_i = 0}$ and there is no value gained from not being served. For a served passenger~${i \in A^*}$, the utility is~${u_i = v_i - p_i}$, the private value $v_i$ for being served minus the price paid $p_i$. Thus, every passenger wants to be served with the smallest price possible. As $v_i$ is private information, we ask every passenger $i$ to submit bid $b_i$ as a proxy for this value. Unfortunately, passenger $i$ can misreport and submit a bid $b_i$ which is different from $v_i$ to maximize their utility.

An auction is \emph{strategy-proof} if every passenger can maximize their utility by bidding ${b_i = v_i}$. Therefore, a single passenger cannot manipulate the auction to their advantage by misreporting their true value. Since we restrict our attention to auctions that are strategy-proof we refer to $b_i$ or $v_i$ interchangeably.

Let $\mathcal{A}$ be a set of trips, or \emph{alternatives}, from which the winning trip is chosen. The \emph{social welfare} of a trip is the sum of the values of the served passengers minus the cost for the driver, that is, $V(A) = \sum_{i \in A} v_i - \mathrm{\emph{cost}}(A)$. An auction \emph{maximizes social welfare} if~$V(A^*) = \max \{V(A) |\, A \in \mathcal{A} \}$. Thus, such an auction can find which trip provides the best value to society.

Another important property that an auction for ridesharing systems can have is to guarantee that the driver always covers their cost. An auction is said to be (weakly) \emph{budget-balanced} if the total price paid by the passengers is at least the cost for the driver\footnote{The word weakly is used in contrast with the concept of a strong budget balance, where the price paid by the passengers is exactly the driver's cost.}, that~is, $\sum_{i \in I} p_i \geq \mathrm{\emph{cost}}(A^*)$.

Finally, \emph{profit} is an important goal both for drivers and for the organization running the system and is defined as~$\mathit{profit}(A^*) = \sum_{i \in I} p_i - \mathrm{\emph{cost}}(A^*)$. To \emph{maximize profit} is to maximize the surplus payment after paying the driver~cost.

\section{Auctions}\label{sec:auctions}

In this section, we propose three auctions that are strategy-proof and budget-balanced. For later comparison, we also present a previously existing version of VCG with reserve prices, denoted by $\mathrm{VCG}_r$, that has both properties~\cite{HartlineR09}. We start by presenting an unweighted version of our main auction to introduce the notation and main ideas. Afterward, we present our main auction, proving that it is strategy-proof and budget-balanced. Then, we present adaptations of the classical VCG auction to our setting and conclude with a discussion on two methods to choose the passengers' reserve prices, an important concept of our~auctions.
\subsection{Unweighted Minimum Surplus auction}

We present a simpler auction, called the \emph{Unweighted Minimum Surplus} (UMS) auction, which will be the starting point for the weighted version.

For the auction to be budget-balanced, it must ensure that the cost $\mathrm{\emph{cost}}(A^*)$ of the winning trip will be covered by the prices paid. There is no direct way to divide~$\mathrm{\emph{cost}}(A^*)$ as $\mathrm{\emph{cost}}$ is not an additive function. Nonetheless, we must decide on how to allocate the trip cost among the winners. To achieve this, we define, for each passenger~$i$, a value $r_i$ which estimates the cost incurred on the driver if~$i$ is served. We say that $r_i$ is the \emph{passenger reserve price} of $i$ and the exact way to determine~$r_i$ is a choice to be made by the auction designer. For example,~$r_i$ can be the cost of going from passenger's $i$ pick-up location to their drop-off location (the \emph{direct cost}, discussed later on).

A trip $A$ will be considered \emph{feasible} if the passenger reserve prices are sufficient to cover the cost of the trip and each passenger can pay their reserve price, that is~${\sum_{i \in A} r_i \geq \mathrm{\emph{cost}}(A)}$ and~${v_i \geq r_i}$ for all $i \in A$. The set of alternatives $\mathcal{A}$ for the auction is composed only of feasible trips.

The \emph{surplus} $s_i = v_i - r_i$ of passenger $i$ is the difference between the passenger's bid and the passenger's reserve price. Only passengers with $s_i \geq 0$ are considered in our auction.
Let the \emph{minimum surplus} of a trip $A$ be defined as ${s_{\min}(A) = \min_{i \in A} s_i}$. In the UMS auction, the winning trip $A^*$ is given by the allocation function~${A^* = \arg\max s_{\min}(A)}$ where~$A \in \mathcal{A}$. Ties are broken by some total ordering of $\mathcal{A}$.

We see that in our model the passenger expresses the private value in a single value~$v_i$ for winning, since the value for losing is always zero and the result for other passengers does not matter.
Thus, the UMS auction has a \emph{single-parameter domain} (for discussion on mechanisms with a single-parameter domain, see Nisan et al.~\cite{nisan_algorithmic_2007}). Here, we use an important concept of single-parameter domain mechanisms to define the price to be paid by the passengers.

The \emph{critical value} $crit_i(v_{-i})$, given a vector of values $v_{-i}$ by other passengers and an allocation function $A^*$, is such that for any $v_i$ if~${v_i > crit_i(v_{-i})}$, then ${i \in A^*}$ and if~${v_i < crit_i(v_{-i})}$, then ${i \not\in A^*}$. Thus, the critical value is the threshold for a guaranteed~win.

An allocation function $A^*$ is \emph{allocation monotone} if a passenger that wins when reporting a value $v_i$ also wins when reporting any higher value (when considering $\mathcal{A}$ and all other bids are held constant). Therefore, a passenger cannot go from winning to losing by raising their bid.

To show strategy-proofness, we can use the characterization stated in Theorem~\ref{characterization}. Proof of this characterization can be found in Theorem 9.36 of Nisan et al.~\cite{nisan_algorithmic_2007}.

\begin{theorem}[Characterization of strategy-proofness for single-parameter domains]\label{characterization}
An auction for single-parameter domains is strategy-proof if $A^*$ is allocation monotone and~$p_i$ is such that~${p_i = crit_i(v_{-i})}$ if~${i \in A^*}$ and $p_i = 0$ if $i \not\in A^*$. If~${i \in A}$ for all $A \in \mathcal{A}$, then $p_i$ is a constant independent of the bid of~$i$. \qed
\end{theorem}

It is easy to see that, for the UMS auction, $A^*$ is allocation monotone. Thus, we must only correctly define the price to be paid by winners as the critical value.

We define the \emph{second minimum surplus} of $i$, denoted by $\mathrm{\emph{ss}}_i$ as the largest minimum surplus of a trip $A$ such that $i \notin A$ (and $0$ if no such trip exists). That~is, 
\[\mathrm{\emph{ss}}_i = \begin{cases}
    \max \{ s_{\min}(A) \; | \; A \in \mathcal{A}, i \not\in A\}, & \text{if there is } A\in \mathcal{A} \text{ such that } i \notin A\\
    0, & \text{otherwise,}
\end{cases}\]
and we set the price for passenger $i$ as~${p_i = r_i + \mathrm{\emph{ss}}_i}$ if $i \in A^*$ and $p_i = 0$ otherwise.

With this price definition, one can show that the UMS auction is strategy-proof (by showing that $p_i$ is the critical value). We omit this proof since it is very similar to the one presented for the Weighted Minimum Surplus auction later on.

Finally, notice that the UMS auction is budget-balanced. In fact, the auction chooses a feasible trip $A^*$ and, thus, $\sum_{i \in A^*} r_i \geq \mathrm{\emph{cost}}(A^*)$ and $\mathrm{\emph{ss}}_i \geq 0$ for every $i \in A^*$, from where we conclude that
$\sum_{i \in A^*} p_i \geq \sum_{i \in A^*} r_i \geq \mathrm{\emph{cost}}(A^*)$.

\subsection{Weighted Minimum Surplus auction}

We now propose the \emph{Weighted Minimum Surplus} (WMS) auction for a single-driver, multiple passenger setting, which is a budget-balanced and strategy-proof auction.

We modify the UMS auction to consider the number of passengers on the winning trip. Thus, the \emph{weighted minimum surplus} of $A$ is~$wm(A) = |A| s_{\min}(A)$ which can be interpreted as the potential profit from the trip, considering that all passengers on the trip pay the same amount over their surplus. The winning trip $A^*$ is given by the allocation function~$A^* = \arg\max wm(A)$ where~$A \in \mathcal{A}$. As before, ties are broken by some criteria that do not depend on bids.

Notice that we still have a monotone allocation function, as we show next.

\begin{lemma}
    $A^* = \arg\max wm(A)$ is allocation monotone if ties are broken by a total ordering of $\mathcal{A}$.\label{lemma:monotone}
\end{lemma}

\begin{prf}
    Let $i$ be such that $i \in A^*$. Consider $v'$ such that $v_j' = v_j$ for every $j \neq i$ and ${v'_i > v_i}$. Let $wm'(A)$ denote the weighted minimum surplus of a trip~$A$ when considering~$v'$ instead of $v$. We must show that $i$ still wins under $v'$.

    Suppose, by contradiction, that a set $A$ wins under $v'$ and ${i \not\in A}$. Therefore, ${wm'(A^*) \leq wm'(A)}$. 
    As ${i \not\in A}$, ${wm'(A) = wm(A)}$ and, since $A^*$ wins under $v$, ${wm(A) \leq wm(A^*)}$. 
    Finally, as ${i \in A^*}$, we obtain that ${wm(A^*) \leq wm'(A^*)}$. 
    Chaining the found inequalities, we conclude that ${wm'(A^*) \leq wm'(A) = wm(A) \leq wm(A^*) \leq wm'(A^*)}$.
    But, as ${wm'(A) = wm'(A^*)}$ and ${wm(A) = wm(A^*)}$, we cannot have $A^*$ and $A$ winning under $v$ and under $v'$, respectively, as we use a total ordering of $\mathcal{A}$ for tie breaking.\qed
\end{prf}

Again, we consider Theorem~\ref{characterization} to define the prices that lead to strategy-proofness. Unfortunately, changing the allocation function to consider the size of the trip will lead to more complex critical values. In~\ref{appendix:ums}, we show a class of instances for which the WMS auction is theoretically superior to the UMS auction. This justifies the intuition that taking the cardinality of the trips into account should give better results, even though it leads to a more complicated pricing scheme.

In order to define $p_i$, first we must define the value $wm^*_{i}$ and trip $A'_i$.

Given a passenger~$i$, $wm^*_{i}$ is the largest weighted minimum surplus of trips that do not contain $i$, taking a similar role than $\mathrm{\emph{ss}}_i$ in the UMS auction. Formally,
\[\mathrm{\emph{wm}}^*_i = \begin{cases}
    \max \{ wm(A) \; | \; A \in \mathcal{A}, i \not\in A\}, & \text{if there is } A\in \mathcal{A} \text{ such that } i \notin A\\
    0, & \text{otherwise.}
\end{cases}\]

We also need to define trip $A'_i$. Informally, if $i \in A^*$, then trip $A'_i$ is simply the trip $A \in \mathcal{A}$ of maximum cardinality such that~$i \in A$ and $wm^*_{i} \leq wm(A)$ (a precise definition of $A'_i$ is given afterward).

Given $wm^*_{i}$ and $A'_i$, the price paid by $i$ is defined as
\begin{equation}
    \label{p_i}
    p_i = r_i + \frac{wm^*_{i}}{|A'_i|},
\end{equation}
if $i \in A^*$ and $p_i = 0$ otherwise. The term $r_i$ is sufficient to pay the driver's cost $\mathrm{\emph{cost}}(A^*)$ while the value of the second term (which is non-negative) may be considered entirely as profit to be shared between the driver and the organization that manages the system. Thus, as it happened to the UMS auction, the WMS auction is also budget-balanced.

To prove that $p_i$ is, indeed, the critical value $crit_i(v_{-i})$ and, thus, that the WMS auction is strategy-proof, we need a general definition of $A'_i$.

Formally, $A'_i$ is defined for any $i$ as the set of maximum cardinality among the sets $A \in \mathcal{A}$ such that $i \in A$, $|A| > 1$ and
\begin{equation}\label{A_i}
    wm^*_{i} \leq |A| s_{\min}(A \setminus \{i \}).
\end{equation}
In case of an equality in Equation~(\ref{A_i}), $A'_i$ must also be such that it wins a tie against any~$A$ for which~${wm^*_{i} = wm(A)}$. If no such $A'_i$ that fulfills all the requirements exists, then if~$\{ i \} \in \mathcal{A}$ let~$A'_i = \{ i \}$.

In~\ref{appendix:wms}, we prove that if there is $v_i \in \mathbb{R}$ such that $i\in A^*$ (that is, if there is a high enough bid such that $i$ can be served), then, even when considering this more general definition, $A'_i$ exists and $p_i$ is well-defined.

We now prove that, in the WMS auction, $p_i = crit_i(v_{-i})$ for $i \in A^*$ and, therefore, winners pay the critical value.

First, Lemma~\ref{lemma:vi_smaller} shows that if a winning passenger $i$ lowers their bid below~$p_i$, then they will no longer win.

\begin{lemma}\label{lemma:vi_smaller}
    Let $i$ be a passenger such that $A'_i$ exists and $v_i \in \mathbb{R}$. If $v_i < r_i + \frac{wm^*_{i}}{|A'_i|}$, then $i \not\in A^*$.
\end{lemma}
\begin{prf}
    Assume the existence of $A$ such that $i \in A$ and $wm^*_{i} \leq wm(A)$.  If $A = \{ i \}$, then $|A| \leq |A'_i|$.
    Otherwise, as $wm(A) \leq |A| s_{\min}(A \setminus \{i \})$ and ${wm^*_{i} \leq wm(A)}$, we have $wm^*_{i} \leq |A| s_{\min}(A \setminus \{i \})$ and, from the definition of~$A'_i$ in Equation~(\ref{A_i}), we again obtain that ${|A| \leq |A'_i|}$.

    Combining the fact that ${|A| \leq |A'_i|}$ with the fact that $s_{\min}(A) \leq s_i$ (as $i \in A$), we get~${|A|s_{\min}(A) \leq s_i|A'_i|}$ and applying the hypothesis $v_i - r_i < \frac{wm^*_{i}}{|A'_i|}$ we reach that ${wm(A) < wm^*_{i}}$ which is a contradiction. Therefore, such $A$ cannot exist and any trip containing $i$ loses to the trip of weighted minimum surplus $wm^*_{i}$, so $i$ cannot win if $v_i < r_i + \frac{wm^*_{i}}{|A'_i|}$.\qed
\end{prf}

The second part of the critical value proof is in Lemma~\ref{lemma:vi_larger}. One of its consequences is that if a passenger that is not winning raises their bid above the price they would pay if they were on the winning trip, then they will be on the new winning trip.

\begin{lemma}
    Let $i$ be a passenger such that $A'_i$ exists and $v_i \in \mathbb{R}$. If $v_i > r_i + \frac{wm^*_{i}}{|A'_i|}$, then $i \in A^*$.\label{lemma:vi_larger}
\end{lemma}
\begin{prf}
    We must show if $s_i|A'_i| = (v_i - r_i)|A'_i| > wm^*_{i}$ then $i \in \arg\max wm(A)$. Since $i \in A'_i$ and $A'_i$, by definition, wins in a tie against any trip of value~$wm^*_{i}$, it is sufficient to show that $wm(A'_i) \geq wm^*_{i}$ since~$wm^*_{i}$ is the value of the best trip that does not contain $i$. If $s_i =s_{\min}(A'_i)$, then using the hypothesis that~${s_i |A'_i| > wm^*_{i}}$ we conclude that ${wm(A'_i) > wm^*_{i}}$. Otherwise, if~${s_i \not= s_{\min}(A'_i)}$ then we know that~${A'_i \not= \{ i \}}$ and that $s_{\min}(A'_i \setminus \{ i \})= s_{\min}(A'_i)$. So, by Equation (\ref{A_i}), we have~${wm(A'_i) \geq wm^*_{i}}$. In both cases~$A'_i$ is better than any trip with value $wm^*_{i}$ which is the value of the best trip without~$i$. From where we conclude that $i \in A^*$.\qed
\end{prf}

Finally, we can satisfy the characterization by showing that the auction is allocation monotone and also charges the critical value from winners. In Theorem~\ref{theorem:strategy} we reach the key property of strategy-proofness.

\begin{theorem}
    The WMS auction is strategy-proof.\label{theorem:strategy}
\end{theorem}
\begin{prf}
    By Lemma~\ref{lemma:monotone} we have that $A^* = \arg\max wm(A)$, as defined for the WMS auction, is allocation monotone.  Thus, we only have to show that any passenger $i$ pays the critical value.

    First, notice that if $A'_i$ exists, then $crit_i(v_{-i}) = r_i + \frac{wm^*_{i}}{|A'_i|}$ is the critical value. In fact, from Lemmas~\ref{lemma:vi_smaller} and~\ref{lemma:vi_larger}, we have that for any $v_i$ if~${r_i + \frac{wm^*_{i}}{|A'_i|}} < v_i$ then~$i \in A^*$ and if~$v_i < {r_i + \frac{wm^*_{i}}{|A'_i|}}$ then~${i \not\in A^*}$. Therefore, $crit_i(v_{-i}) = r_i + \frac{wm^*_{i}}{|A'_i|}$ is the critical value for~${A^* = \arg\max wm(A)}$.

    Finally, if $A'_i$ does not exist then, for any $v_i \in \mathbb{R}$, $i \not\in A^*$ and, by the definition of~$p_i$, if $i \not\in A^*$, then $p_i = 0$.

    We conclude that the WMS auction satisfies the conditions of Theorem~\ref{characterization} and is strategy-proof.\qed
\end{prf}

\subsubsection{Example}

\begin{figure}[h]
    \centering
    \includegraphics[scale=0.3]{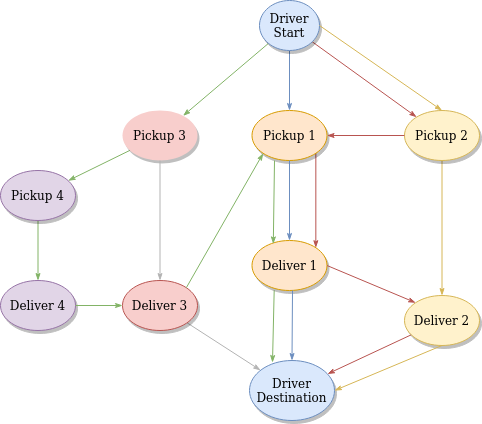}
    \caption{Nodes and edges of example instance. All edges have unit cost.\label{fig:graph}}
\end{figure}

Here we illustrate an instance of the Weighted Minimum Surplus auction. The input graph consisting of a driver with a start and end position and four passengers is presented in Figure~\ref{fig:graph}.

The trips with highlighted paths are ${R = \{ 1, 2 \}}$ in red, ${G = \{ 1, 3, 4 \}}$ in green, ${B = \{ 1 \}}$ in blue and~${Y = \{ 2 \}}$ in yellow with multiple arrows between two nodes indicating an edge being used by multiple highlighted trips. Table~\ref{table:inputs} shows the input bids, passenger reserve prices, and trip costs. We can verify that these four trips are feasible since~$\sum_{i \in A} r_i \geq \mathrm{\emph{cost}}(A)$ for all of $R$, $B$, $G$ and $Y$. Other feasible trips are omitted because they do not affect the outcome of the auction, so without loss of generality, the set of trips is~$\mathcal{A} = \{ R, G, B, Y \}$.

The winning trip is $A^* = R$ since $R = \arg \max wm(A)$ for $A \in \mathcal{A}$, so the two passengers $1$ and~$2$ are served, and their price must be determined. First, we calculate~$p_1$ which depends on~$wm^*_{1}$ and~$A'_1$. Since $Y$ is the only trip that does not contain $1$, trivially $wm^*_{1} = wm(Y)$. By definition~$A'_1$ is trip of maximum cardinality such that $1 \in A$ and $wm^*_{1} \leq wm(A)$ therefore $A'_1 = G$. So we have~${p_1 = r_1 + \frac{wm(Y)}{|G|} = 4 + \frac{8}{3}}$. For~$p_2$ we need~$wm^*_{2}$ and~$A'_2$. The second-highest weighted minimum surplus is that of $G$ and~${2 \not\in G}$, therefore $wm^*_{2} = wm(G)$. The only trip such that $2 \in A$ and $wm(G) \leq wm(A)$ is~$R$ itself, therefore $A'_2 = R$. Finally, we have $p_2 = r_2 + \frac{wm(G)}{|R|} = 10$. In this example, it is interesting to note that even though passenger $1$ had a higher bid than passenger~$2$ and both had the same reserve price, the price $p_2$ was higher.

\begin{table}\scriptsize
    \centering
    \caption{Inputs to the example.\label{table:inputs}}
    \subfloat[Passenger value, reserve price and surplus.]{
        \begin{tabular}{cccc}
            \toprule
            Passenger & $v_i$ & $r_i$ & $s_i = v_i - r_i$ \\
            \midrule
            1         & 14    & 4     & 10                \\
            2         & 12    & 4     & 8                 \\
            3         & 8     & 4     & 4                 \\
            4         & 10    & 6     & 4                 \\
            \bottomrule
        \end{tabular}
    }
    \qquad
    \subfloat[Trip cost, size and weighted minimum surplus.]{
        \begin{tabular}{cccccc}
            \toprule
            Trip & Passengers      & $\mathrm{\emph{cost}}(A)$ & $|A|$ & $s_{\min}(A)$ & $wm(A)$ \\
            \midrule
            R    & $\{ 1, 2 \}$    & 5          & 2     & 8             & 16      \\
            G    & $\{ 1, 3, 4 \}$ & 7          & 3     & 4             & 12      \\
            B    & $\{ 1 \}$       & 3          & 1     & 10            & 10      \\
            Y    & $\{ 2 \}$       & 3          & 1     & 8             & 8       \\
            \bottomrule
        \end{tabular}
    }
\end{table}

\subsection{VCG-based auctions}

Social welfare, which is the combined economic gains and losses for all participants, is commonly used to measure the economic efficiency of an auction. The Vickrey-Clarke-Groves (VCG)~\cite{vickrey_counterspeculation_1961, clarke1971multipart, groves_incentives_1973} mechanism is a widely studied general method to obtain a strategy-proof auction that maximizes social welfare.

For the ridesharing problem, there are at least two ways we can define the social welfare of a trip. If the welfare of the driver is not considered, it is simply ${V(A) = \sum_{i \in A} v_i}$. If we consider the driver to be a participant with value $-\mathrm{\emph{cost}}(A)$, taking into account the social impact of longer trips, then the welfare is defined as $V(A) = \sum_{i \in A} v_i - \mathrm{\emph{cost}}(A)$. This choice does not impact the theoretical properties of the auction. Given a definition for $V(A)$, let~$V_{-i}(A) = V(A) - v_i$ if~$i \in A$ and~${V_{-i}(A) = V(A)}$ otherwise. We then obtain a VCG mechanism, using Clarke's pivot rule, by defining the winner as $A^* = \arg\max_{A \in \mathcal{A}} V(A)$ and the price for $i$ as~${p_i = \max_{A \in \mathcal{A}} \{ V_{-i}(A)\} - V_{-i}(A^*)}$. The economic interpretation is that $i$ is being charged their externality, or the welfare they take away from the rest of the society by participating.

The VCG mechanism is strategy-proof and maximizes social welfare by definition, but it is not budget-balanced in this case. As budget-balance is a crucial requirement to incentivize drivers in the ridesharing problem, we can consider two options: use the \emph{surplus welfare} as a measure of welfare that is suitable for budget-balanced auctions or use reserve prices. Both options lead to two different auctions that we present next.

The surplus welfare of an alternative~$A$ is defined as $V_{s}(A) = \sum_{i \in A} s_i$. The surplus welfare of an auction is that of the winning trip $V_{s}(A^*)$. With this in mind, we now propose the $\mathrm{VCG}_s$ auction, which maximizes the surplus welfare instead of the social welfare and achieves budget balance. As in~UMS and WMS, we restrict~$A^*$ to feasible trips such that for all~$A \in \mathcal{A}$ we have~${\sum_{i \in A} r_i \geq \mathrm{\emph{cost}}(A^*)}$ and $v_i \geq r_i$ for all~$i \in A$. Since the sum of passenger reserve prices is always greater or equal to the cost, conversely the surplus welfare will always be less or equal to the social welfare. So the nearer~the passenger reserve prices are to the cost, the nearer the surplus welfare will be to the social welfare. The winning trip and price definitions are the same as the ones for the~VCG auction, but with the surplus welfare in place of the social welfare and the surplus in the place of the bid. The winning trip is~$A^* = \arg\max_{A \in \mathcal{A}} V_{s}(A)$, and the price is~$p_i = r_i + \max_{A \in \mathcal{A}} \{ \sum_{j \in A, i \neq j} s_j\} - \sum_{j \in A^*, i \neq j} s_j$ for~$i \in A^*$, and~$p_i = 0$ for~$i \not\in A^*$.

The $\mathrm{VCG}_s$ auction is budget-balanced since \[\max_{A \in \mathcal{A}} \left\{ \sum_{j \in A, i \neq j} s_j\right\} - \sum_{j \in A^*, i \neq j} s_j \geq 0.\]
It is also strategy-proof, 
which can be easily proven using known results for VCG auctions~\cite{nisan_algorithmic_2007}.

As a mean of comparison, in our experiments, we also consider the $\mathrm{VCG}_r$ auction~\cite{HartlineR09}, a VCG auction with bidder reserve prices, which is budget-balanced and strategy-proof. In this auction, the social welfare of an alternative $A$ is, as in the VCG auction, $V(A) = \sum_{i \in A} v_i - \mathrm{\emph{cost}}(A)$ and the winning alternative is again ${A^* = \arg\max_{A \in \mathcal{A}} V(A)}$. Nonetheless, there are three differences when compared with~VCG:\@ first, only bidders with $v_i \geq r_i$ participates in the auction; second, only alternatives with $\mathrm{\emph{cost}}(A) \leq \sum_{i \in A} r_i$ are considered; and, third, winners pay \[p_i = \max\{\max_{A \in \mathcal{A}} \{ V_{-i}(A)\} - V_{-i}(A^*), r_i\},\] that is, $r_i$ is the reserve price of bidder $i$. Thus, every winning bidder $i$ pays at least their reserve price $r_i$, the price paid is at most $v_i$ and the winning alternative has a cost at most the sum of the price paid by the winning bidders.

\subsection{Setting Passengers' Reserve Price}\label{section:setting_cost}

For a driver to be willing to participate in the system, the price of the trip should at least cover the cost to serve the winning trip. The exact cost for a driver to serve a trip is difficult to model, since the costs perceived by the driver may have complex social and psychological aspects~\cite{lee2015working}. Usually, it is sufficient for the driver to achieve a lower cost than the one they would have by traveling~alone.

This section explores some ways in which the reserve price $r_i$ of passenger~$i$ can be defined in the auctions previously presented. We propose two methods of defining the passengers' reserve price based on passengers' pick-up and drop-off locations: the \emph{direct cost} and the \emph{upper bound cost}. Developing other reserve price functions is an interesting research question.
\paragraph{Direct cost}
The direct cost is to make $r_i$ equal to the cost incurred on the driver from traveling between the pick-up and drop-off location of passenger $i$. Notice that this value is usually much lower than the real cost that the passenger would have to go from his pick-up location to his drop-off location using other means of transportation, such as taxis. 
Also, even using their car (which is unrealistic to assume that every passenger will have such possibility) could lead to a greater cost due to parking costs.
Thus, it is expected that several passengers will have $v_i \geq r_i$. As any feasible trip $A$ must satisfy ${\sum_{i \in A} r_i \geq cost(A)}$, this method of defining passengers reserve prices has the interesting property that any feasible trip will result in a global cost saving in comparison to each passenger going from their pick-up to drop-off point by themselves.

\paragraph{Upper bound cost}
The upper bound cost is to make $r_i$ equal to the round trip cost of passenger $i$. The round trip cost is the minimum cost for the driver to go from the starting location (or destination) to pick up and drop-off passenger $i$ and then return to the starting location (or destination). When the distances respect the triangle inequality, this definition of reserve price makes the restriction that~$v_i \geq r_i$ for all~$i \in A$ necessary and sufficient for $A$ to be a feasible trip, as it has the property that~$\sum_{i \in P} r_i \geq \mathrm{\emph{cost}}(A)$ is satisfied for any $A$ because the cost of making all the round trips in a sequence is an upper bound to the optimal cost of the trip. Note that this reserve price can still be lower than the cost of other means of transportation, such as a taxi, especially if the driver is near the passenger.

\section{Lower bounds on Surplus Welfare and Profit}\label{sec:bounds}

In this section, we present worst-case guarantees for the surplus welfare and the surplus profit obtained by the WMS auction when the set $\mathcal{A}$ is \emph{downward closed}, that is, for all $A \in \mathcal{A}$ and for all $B \subseteq A$ we have~$B \in \mathcal{A}$.

\subsection{Lower bound on the surplus welfare of the WMS auction}

Given a family of trips $\mathcal{A}$, we define the \emph{surplus welfare ratio} of an auction as the ratio between the maximum surplus welfare and the obtained surplus welfare:
\[
    \frac{\max_{A \in \mathcal{A}} \{ V_{s}(A) \}}{V_{s}(A^*)},
\]
where $A^*$ is the winning trip among $\mathcal{A}$. The $\mathrm{VCG}_s$ auction by construction has a surplus welfare ratio of~$1$ for any $\mathcal{A}$, which is the minimum possible welfare ratio and, therefore, optimal. We will analyze the surplus welfare ratio of the WMS auction. First, we show a negative result for non-downward closed alternatives.

\begin{theorem}
    With non-downward closed alternatives, $\mathcal{A}$ can be constructed such that the WMS auction has an arbitrarily large surplus welfare ratio.
\end{theorem}
\begin{prf}
    We show that if the alternatives are not downward closed, then, given a parameter $n \in \mathbb{R}$, a set of alternatives may be constructed such that the surplus welfare ratio of the WMS auction is proportional to $n$. That is, the surplus welfare ratio can be made arbitrarily large.

    Let the feasible alternatives be $\mathcal{A} = \{ A, B \}$ where $A = \{ n, 1 \}$ and $ B = \{ 3 \}$. The elements of the alternatives denote the surplus $s_i$ of the contained passengers, this is without loss of generality since any other information is irrelevant. We see that $wm(A) \leq 2$ regardless of $n$ while $wm(B) = 3$. Therefore, ${A^* = B}$ and $V_{s}(A^*) = 3$. However, $V_{s}(A) = n + 1$, so we have a surplus welfare ratio of~$\frac{n + 1}{3}$.\qed
\end{prf}

With that, we turn to the case of downward closed alternatives. This allows the WMS auction to obtain a surplus welfare ratio of $H_{\max |A|}$, which is the harmonic number of the maximum cardinality found in~$\mathcal{A}$. Note that this is only meaningful if a bound on $\max |A|$ can be given. In our ridesharing problem, the size of the alternatives is bounded by the maximum capacity of each vehicle, for example, if cars with a maximum capacity of $3$ passengers are used, this would give a bound of $H_3 = \frac{11}{6} \approx 1.83$.

\begin{theorem}
    Given a downward closed family of alternatives~$\mathcal{A}$, the WMS auction has a surplus welfare ratio of at most $H_{\max |A|}$ for $A \in \mathcal{A}$.\label{theo:approx}
\end{theorem}

\begin{prf}
    Without loss of generality, assume an $\mathcal{A}$ and then some $A \in \mathcal{A}$. Let $(1, \dots, i, \dots, |A|)$ be an ordering of the passengers in $A$ such that $v_i \geq v_{i + 1}$ for~$i \leq |A|$. Let $A_i$ be the alternative consisting of passengers $1$ to $i$. Since $A_i \subseteq A$ and the alternatives are downward closed we have $A_i \in \mathcal{A}$, and, in the WMS auction, we have that~${wm(A^*) = \max_{A \in \mathcal{A}} \{ wm(A) \}}$. Thus, for all $i$ we get that $wm(A_i) \leq wm(A^*)$ which gives $s_{\min}(A_i) \leq \frac{wm(A^*)}{|A_i|}$. By construction,~${s_i = s_{\min}(A_i)}$ and~$|A_i| = i$, so $s_i \leq \frac{wm(A^*)}{i}$. The surplus welfare of alternative $A$ is defined as ${V_{s}(A) = \sum_{i \in A} s_i}$, so we have obtained ${V_{s}(A) \leq \sum_{i = 1}^{|A|} \frac{wm(A^*)}{i}}$, rewriting the right-hand side as~${wm(A^*) \sum_{i = 1}^{|A|} \frac{1}{i}}$ we find that the surplus welfare of $A$ is bounded by $V_{s}(A) \leq wm(A^*)H_{|A|}$. Loosening the bound we have~${V_{s}(A) \leq wm(A^*)H_{\max |A|}}$ for all $A \in \mathcal{A}$. Noting the fact that~${|A|\min_{i \in A} s_i \leq \sum_{i \in A} s_i}$ therefore~${wm(A) \leq V_{s}(A)}$ for any $A$ and in particular~${wm(A^*) \leq V_{s}(A^*)}$ we may further loosen the inequality to ${V_{s}(A) \leq V_{s}(A^*)H_{\max |A|}}$ for all~$A \in \mathcal{A}$. It is now immediate that
    \[
        \frac{\max_{A \in \mathcal{A}} \{ V_{s}(A) \}}{V_{s}(A^*)} \leq H_{\max |A|}. \;
    \]
    Therefore, the surplus welfare ratio of the WMS auction is at most~$H_{\max |A|}$.\qed{}
\end{prf}

In~\ref{appendix:wms}, we show that the bound given in Theorem~\ref{theo:approx} on the surplus welfare ratio is tight.

It is worth noting that Theorem~1 of Goel and Khani~\cite{goel2014revenue}, which was independently developed for application in advertisement auctions, has similarity with Theorem~\ref{theo:approx}, also bounding the worst-case welfare by a harmonic number in a scenario with downward closed alternatives.

We may ask whether the downward closed alternative condition applies to the ridesharing problem. An instance of the ridesharing problem may not be downward closed due to the restriction that all alternatives $A \in \mathcal{A}$ are feasible with~${\sum_{i \in A} r_i \geq \mathrm{\emph{cost}}(A)}$. For example, let~${A = \{ 1 \}}$ and $B = \{ 1, 2 \}$ with $r_1 = 1$ and $r_2 = 2$. Then $\mathrm{\emph{cost}}(A) = 2$ and~$A$ is not feasible ($A \not\in \mathcal{A}$), but since $\mathrm{\emph{cost}}(B) = 3$ then $r_1 + r_2 \geq \mathrm{\emph{cost}}(B)$ and $B \in \mathcal{A}$. Therefore, $\mathcal{A}$ is not downward closed and Theorem~\ref{theo:approx} does not necessarily apply to all ridesharing instances.

Nonetheless, the \emph{upper bound cost} presented in Section~\ref{section:setting_cost} when distances respect the triangle inequality
produces a set~$\mathcal{A}$ of trips which is downward closed as $cost(A)$ is upper bounded by the sum of the round trip costs of the passengers in $A$.

\subsection{Lower bound on the surplus profit of the WMS auction}

Analogous to the surplus welfare, we can define the surplus profit as $\mathit{profit}_{s}(A^*) = \sum_{i \in A^*} p_i - r_i$. Recalling that the total profit is $\mathit{profit}(A^*) = \sum_{i \in A^*} p_i - \mathrm{\emph{cost}}(A^*)$ and that $\sum_{i \in A^*} r_i \geq \mathrm{\emph{cost}}(A^*)$ we see that~${\mathit{profit}_{s}(A^*) \leq \mathit{profit}(A^*)}$, so the surplus profit is a lower bound to the total profit. Under certain assumptions, we show that the surplus profit of the Weighted Minimum Surplus auction is guaranteed to be at least $\frac{1}{\max_{A \in A} \{ |A| \}}$ of the obtained surplus welfare.

\begin{theorem}
    Assume that $|A^*| \geq 2$ and that for all $i \in A^*$ it is true that $A^* \setminus \{ i \} \in \mathcal{A}$. Then in the WMS auction $\mathit{profit}_{s}(A^*) \geq \frac{V_{s}(A^*)}{\max_{A \in A} \{|A| \}}$.\label{theo:profit}
\end{theorem}
\begin{prf}
    By the definitions of $\mathit{profit}_{s}(A^*)$ and $p_i$, we have that in the WMS auction ${\mathit{profit}_{s}(A^*) = \sum_{i \in A^*} \frac{wm^*_{i}}{|A'_i|}}$. By the assumption that $|A^*| \geq 2$, we know that $A^* \setminus \{ i \}$ is not empty and since ${A^* \setminus \{ i \} \subset A^*}$ we have~${s_{\min}(A^* \setminus \{ i \}) \geq s_{\min}(A^*)}$. By hypothesis $A^* \setminus \{ i \} \in \mathcal{A}$, so ${wm^*_{i} \geq (|A^*| - 1)s_{\min}(A^*)}$ for all~$i \in A^*$. Substituting into the definition of surplus profit, we have
    \[{\mathit{profit}_{s}(A^*) \geq \sum_{i \in A^*} \frac{(|A^*| - 1)s_{\min}(A^*)}{|A'_i|}},\] considering that the worst case is when $|A^*| = 2$ and~$|A'_i| = \max_{A \in A} \{ |A| \}$ we~get \[{\mathit{profit}_{s}(A^*) \geq \sum_{i \in A^*} \frac{s_{\min}(A^*)}{\max_{A \in A} \{ |A| \}}},\]finally, by the definition of surplus welfare, we conclude that~$\mathit{profit}_{s}(A^*) \geq \frac{V_{s}(A^*)}{\max_{A \in A} \{ |A| \}}$.\qed
\end{prf}
Combining Theorem~\ref{theo:approx} and~\ref{theo:profit} we obtain Theorem~\ref{theo:ratio} as an immediate consequence.

\begin{theorem}\label{theo:ratio}
    If $\mathcal{A}$ is downward closed and $|A^*| \geq 2$, then
    \[
        \mathit{profit}_{s}(A^*) \geq \frac{\max_{A \in \mathcal{A}} \{ V_{s}(A) \}}{H_{\max |A|}\max_{A \in A} \{ |A| \}}.
    \]
\end{theorem}
\begin{prf}
    If $\mathcal{A}$ is downward closed and $|A^*| \geq 2$ then Theorem~\ref{theo:approx} and~\ref{theo:profit} both apply, and the result is immediate.\qed{}
\end{prf}

It is a remarkable result to achieve a bound on how much of the surplus welfare is converted to surplus profit by the auction, considering that the VCG auction maximizes welfare but may obtain little or no profit~\cite{lonelyvcg}. 
In particular, we present an example where $\mathrm{VCG}_r$ and $\mathrm{VCG}_s$ performs poorly in terms of profit when compared with the WMS auction. Consider a set $I = \{1, \dots,  k\}$ of $k$ passengers such that, for $i \in I$, $r_i = 1$, $v_i = M$ where $M > 1$ and $\mathrm{\emph{cost}}(A) = |A|/2$ for all $A \subseteq I$. Thus, we have downward-closed alternatives and the driver can serve any subset of passengers. All the three auctions ($\mathrm{VCG}_r$, $\mathrm{VCG}_s$ and WMS) will serve all passengers. 
For $i \in I$, in the case of $\mathrm{VCG}_s$, the price $p_i$ paid is 
\[r_i + \max_{A \in \mathcal{A}} \left\{ \sum_{j \in A, i \neq j} s_j\right\} - \sum_{j \in A^*, i \neq j} s_j
=
1 + (k - 1)(M - 1) - (k - 1)(M - 1) = 1,
\] and in the case of $\mathrm{VCG}_r$, the price $p_i$ paid is
\begin{align*}
    & \max\{\max_{A \in \mathcal{A}} \{ V_{-i}(A)\} - V_{-i}(A^*), r_i\}\\
    & \qquad = \max\left\{(k - 1)M - \frac{k - 1}2 - \left((k - 1)M  - \frac{k}2\right), 1\right\}\\
    & \qquad = 1.
\end{align*}

That is, in this case, both $\mathrm{VCG}_s$ and $\mathrm{VCG}_r$ charges exactly the cost of serving all the passengers. Now, notice that, for all $A \subseteq I$, $wm(A) = |A|s_{\min}(A) = |A|(M - 1)$. Thus, for all $i \in I$, $wm^*_{i}= (k-1)(M - 1)$ and $A_i' = I$. In this case, we have
\[
    p_i = r_i + \frac{wm^*_{i}}{|A'_i|} = 1 + \frac{(k-1)(M - 1)}k,
\] 
from where we conclude that the profit in the WMS auction is $(k-1)(M - 1)$, a value that can be made arbitrarily large as the number of passengers or the value of the bids increases.

For other auctions that have a lower bound on welfare and profit, see the work on competitive auctions for digital goods by Goldberg et al.~\cite{goldberg2006competitive}.

\section{Computational Experiments}\label{sec:experiments}
In this section, we present computational experiments run to empirically analyze the proposed auctions. In what follows, we present the underlying routing problem considered, algorithms to solve the proposed auctions, and the experiments performed.

\subsection{Instance model}

First, we define our model for the input instance. Each instance is a complete graph in which pick-up nodes belong to the set $P = \{1, \ldots, n\}$ and the drop-off nodes to the set $D = \{n + 1, \ldots, 2n\}$ where~$n$ is the number of passengers. The surplus and reserve price of passenger $i$ are non-negative constants $s_i$ and~$r_i$, respectively. The driver's starting location is node $0$ and the driver's destination is node~${2n + 1}$. Therefore, the set of all nodes in the graph is~${N = P \cup D \cup \{0, 2n + 1\}}$. The maximum capacity of passengers in the vehicle is $Q$ and the change in the number of passengers in the vehicle when visiting node $i$ is $q_i$ such that if $i \in P$ then $q_i = 1$, if $i \in D$ then $q_i = -1$, and if~${i \in \{0, 2n + 1\}}$ then~$q_i = 0$. For the driver to go from node $i$ to node $j$, therefore crossing the edge~$(i, j)$, it takes $t_{ij}$ time and costs $c_{ij}$. The maximum pick-up time and travel time of passenger~$i$ are respectively $k_i$ and $l_i$, and the driver's maximum arrival time is $T$.

\subsection{ILP-based Algorithms}
Now we present an exact algorithm based on integer linear programming (ILP) to solve the WMS auction. There is extensive literature on using integer linear programming to solve vehicle routing problems exactly~\cite{golden2008vehicle,Pessoa20,Vidal20,Cagri20,Mor22,Freitas23}, and our formulation of the constraints is based on the one presented for the Dial-A-Ride Problem (DARP) by Cordeau~\cite{cordeau_branch-and-cut_2003}, simplified for a single driver.

We start describing the ILP model by its variables, whose values will be set to optimize the objective function. Each edge~$(i,j)$ has an associated binary variable $x_{ij} \in \{0, 1\}$ which indicates if the edge belongs to the trip. The real and non-negative variable $B_i$ is the instant in which the driver visits the node~$i$. The real and non-negative variable $Q_i$ is the number of passengers in the vehicle after visiting node~$i$.

The objective function is to maximize $m \cdot s_{\min}$ where the real and non-negative variable $s_{\min}$ is the minimum surplus of the trip and variable $m$ is a non-negative integer, representing the number of passengers in the trip. A non-linear but direct formulation~is
{\small
        \begin{alignat}{4}
             & \omit\rlap{$\max m \cdot s_{\min}$}                                                                                                                                                                   \label{obj}                                                                                                                                                                   \\
            \label{constr:driver_start}
             & \text{subject to}                                                                                                                                                                                                  & \sum_{j \in N} x_{0j}                          & = 1                                                        & \qquad &                                          \\
            \label{constr:driver_end}
             &                                                                                                                                                                                                                    & \sum_{i \in N} x_{i, 2n+1}                     & = 1                                                                                                            \\
            \label{constr:enter_leave}
             &                                                                                                                                                                                                                    & \sum_{j \in N} x_{ji}                          & = \sum_{j \in N} x_{ij}                                    &        & \forall i \in P \cup D                   \\
            \label{constr:pickup_deliver}
             &                                                                                                                                                                                                                    & \sum_{j \in N} x_{ij}                          & = \sum_{j \in N} x_{i+n,j}                                 &        & \forall i \in P                          \\
            \label{constr:set_time}
             &                                                                                                                                                                                                                    & (B_i + t_{ij})x_{ij} \le B_j                   & \le T                                                      &        & \forall i, j \in N                       \\
            \label{constr:max_pickup}
             &                                                                                                                                                                                                                    & B_i                                            & \le k_i                                                    &        & \forall i \in P                          \\
            \label{constr:precedence_and_max_time}
             &                                                                                                                                                                                                                    & 0 \leq B_{n+i} - B_i                           & \leq l_i                                                   &        & \forall i \in P                          \\
            \label{constr:capacity}
             &                                                                                                                                                                                                                    & (Q_i + q_{j})x_{ij} \leq Q_j                   & \leq Q                                                     &        & \forall i, j \in P \cup D                \\
            \label{constr:fixed_cost}
             &                                                                                                                                                                                                                    & \sum_{i, j \in N} c_{ij}x_{ij} - c_{0, 2n + 1} & \leq \sum_{i \in P} \left(r_i \sum_{j \in N} x_{ij}\right)                                                     \\
            \label{constr:n_pass}
             &                                                                                                                                                                                                                    & m                                              & = \sum_{j \in N} x_{ij}                                    &        & \forall i \in P                          \\
            \label{constr:min_surplus}
             &                                                                                                                                                                                                                    & s_{\min}                                       & \leq s_i + M \left(1 - \sum_{j \in N} x_{ij}\right)        &        & \forall i \in P                          \\
             &                                                                                                                                                                                                                    & x_{ij}                                         & \in \{0,1\}                                                &        & \forall i, j \in N\label{constr:domain1} \\
             &                                                                                                                                                                                                                    & B_i                                            & \in \mathbb{R}^+                                                   &        & \forall i \in N\label{constr:domain2}    \\
             &                                                                                                                                                                                                                    & Q_i                                            & \in \mathbb{R}^+                                           &        & \forall i \in N\label{constr:domain3} \\
             &                                                                                                                                                                                                                    & m                                            & \in \mathbb{Z}^+                                           &        &\label{constr:domain4}\\
             &                                                                                                                                                                                                                    & s_{\min}                                            & \in \mathbb{R}^+                                           &        & \label{constr:domain5}
        \end{alignat}
}

Constraints~(\ref{constr:driver_start}) to~(\ref{constr:capacity}) are routing constraints, and Constraints~(\ref{constr:fixed_cost}) to~(\ref{constr:min_surplus}) together with Objective Function~(\ref{obj}), guarantees that the maximum weighted minimum surplus is computed while ensuring that the route's cost is not above the sum of the served passengers reserve prices. Finally, Constraints~(\ref{constr:domain1}) to (\ref{constr:domain5}) are the domain constraints.

We start by explaining Constraints~(\ref{constr:driver_start}) to~(\ref{constr:fixed_cost}).
Constraints (\ref{constr:driver_start}), (\ref{constr:driver_end}) and (\ref{constr:enter_leave}) are flow constraints which ensure that the variables~$x_{ij}$ of value~$1$ define a simple path starting at the driver's current location and ending at the driver's destination.
No cycles are formed as otherwise, Constraint (\ref{constr:set_time}) would make arrival times go to infinity.
Constraint~(\ref{constr:pickup_deliver}) guarantees that passengers are picked up if and only if they are dropped off.
Constraint~(\ref{constr:set_time}) sets the arrival times at each node and bounds them by the driver's maximum arrival time. Time Constraints~(\ref{constr:precedence_and_max_time}) ensure that passengers are picked up before being dropped off, and that maximum travel times are respected, while Constraint~(\ref{constr:max_pickup}) enforces maximum pick-up times. Capacity Constraints~(\ref{constr:capacity}) set the number of passengers in the vehicle at each node and ensure that the maximum capacity is respected.
Constraint~(\ref{constr:fixed_cost}) is the condition that $\sum_{i \in A^*} r_i \geq \mathrm{\emph{cost}}(A^*)$, where~$\mathrm{\emph{cost}}(A)$ is defined as the detour cost, which is the total cost of the route minus the direct distance to the driver's destination. We can generally use any other definition of cost that can be formulated as a linear expression. This constraint and the non-negativity of~$s_i$ ensure that all solutions of the model are feasible trips.
The quadratic Constraints~(\ref{constr:set_time}) and~(\ref{constr:capacity}) may be linearized respectively as $B_i + t_{ij} - M(1 -x_{ij}) \leq B_j$ and~$Q_i + q_j - M(1 -x_{ij}) \leq Q_j$ where~$M$ is a sufficiently large constant.

Finally, we explain Constraints~(\ref{constr:n_pass}) and~(\ref{constr:min_surplus}), commenting on how they, along with Objective Function~(\ref{obj}), lead to the computation of the alternative with maximum weighted minimum surplus.
Constraint~(\ref{constr:n_pass}) fixes $m$ as the number of served passengers. Constraint~(\ref{constr:min_surplus}), where value~$M$ is a sufficiently large constant (for example the largest bid), guarantees that $s_{\min}$ is at most $s_i$ for any $i$ such that $\sum_{j \in N} x_{ij} = 1$. Thus, as we seek to maximize $m \cdot s_{\min}$ and~$m$ is non-negative, in any optimal solution $s_{\min}$ is the minimum $s_i$ over $i$ such that $\sum_{j \in N} x_{ij} = 1$ and, thus, the objective value is $\max |A| s_{\min}(A)$ as desired.

Notice that the objective function is quadratic as written, which can be solved by two means. The first is to introduce a variable $z_i$ for each passenger~$i$ and consider the following ILP:\@
    \begin{alignat}{4}
         & \text{maximize}         &     & \sum_{i \in P} z_i \nonumber                                                                  \\
         & \text{subject to} \quad & z_i & \leq s_i\sum_{j \in N} x_{ij} \quad                            & \forall i \in P \nonumber    \\
         &                         & z_i & \leq s_j + \left(1- \sum_{k \in N} x_{jk}\right)s_i \quad      & \forall i, j \in P \nonumber \\
         &                         &     & (\ref{constr:driver_start})-(\ref{constr:fixed_cost})\nonumber\\
         &                         &     & (\ref{constr:domain1})-(\ref{constr:domain3}).\nonumber
    \end{alignat}
Let $A$ be the set of passengers in a solution. This formulation ensures that, for $i \in A$, we have~${z_i = \min_{j \in A} s_j}$ and, for $i \notin A$, we have $z_i = 0$. Therefore, $\max \sum_{i \in P} z_i  = m \cdot  s_{\min}$ and the linear formulation is equivalent to the original quadratic one. The variables $m$ and $s_{\min}$ are no longer present in the model, but their values can be determined from a solution to this formulation. Another possibility is to consider that the driver can pick up at most $k$ passengers and run the model for each fixed $m \in \{1, \dots, k\}$.

\begin{algorithm}
    \caption{WMS}\label{alg:WMS}
    \scriptsize
    \begin{algorithmic}
        \Procedure{WMS}{$\mathcal{I}$} \Comment{$\mathcal{I}$ denotes the problem's instance}
        \State $A^* \gets \Call{Served}{\mathcal{I}}$ \Comment{Get the set of served passengers}
        \For{$i \notin A^*$}
        \State $p_i \gets 0$ \Comment{Unserved passengers pay $0$}
        \EndFor
        \State $U \gets A^*$ \Comment{$U$ is the set of unpriced passengers}
        \While{$U \neq \emptyset$}
        \State $A \gets \Call{Served'}{\mathcal{I}, U}$ \Comment{Best trip without at least one element of $U$}
        \For{$i \in U \setminus A$}
        \State $wm_i^* \gets wm(A)$ \Comment{$A$ is the best trip without $i$}
        \State $A_i' \gets \Call{Largest}{\mathcal{I}, i, wm_i^*}$ \Comment{Finds $\arg\max (|A| \colon |A|s_{\min}(A) \geq wm_i^* \wedge i \in A)$}
        \State $p_i = wm_i^* \,/\, |A_i'|$
        \EndFor
        \State $U \gets U \setminus A$
        \EndWhile
        \State $route \gets \Call{Route}{\mathcal{I}, A^*}$ \Comment{Compute the optimal route to serve $A^*$}
        \EndProcedure
    \end{algorithmic}
\end{algorithm}

Next, we present Algorithm~\ref{alg:WMS}, which determines the winning trip, the prices to be paid, and the driver's route. First, one must solve an optimization problem to obtain a set $A^*$ of served passengers, which can be done with the presented ILP and is denoted as function \textsc{Served} in the pseudocode. Then, to compute the price paid by each passenger in $A^*$ (as all passengers not in $A^*$ always pay zero), it must solve some optimization problems to compute $wm_i^*$ and $A_i'$ for~${i \in A^*}$.

In order to do so, Algorithm~\ref{alg:WMS} maintains a set $U$ of unpriced passengers. While this set is not empty, we compute a trip $A$ with maximum weighted surplus among those that do not serve at least one passenger of~$U$. In the algorithm, this is represented by a function \textsc{Served'} which also receives set~$U$. This optimization can be done by adding the following constraint to the original~ILP:\@
    \begin{align}
        \sum_{i \in U} \sum_{j \in N} x_{ij} \leq |U| - 1.
    \end{align}
Notice that, by the definition of trip $A$, we have that $wm_i^* = wm(A)$.

Next, the algorithm must compute $A_i'$ for $i \in U \setminus A$, which is denoted by a call to function \textsc{Largest} in the pseudocode. This can be done by changing the ILP objective function to $\max \sum_{i' \in P} \sum_{j \in N} x_{i'j}$ and adding the following constraints:
\begin{align}
    m \cdot s_{\min}      & \geq wm_i^* \\
    \sum_{j \in N} x_{ij} & = 1,
\end{align}
which computes, given $i \in N$ and $wm_i^*$, the largest set $A_i'$ such that $i \in A_i'$ and ${|A_i'|s_{\min}(A_i') \geq wm_i^*}$.

Finally, we compute the optimal driver's route to serve set $A^*$, which is represented by function \textsc{Route} in the algorithm. Using the ILP, this can be done by adding the following constraints:
\begin{align}
    \sum_{j \in N} x_{ij} & = 1 &  & \forall i \in A^*    \\
    \sum_{j \in N} x_{ij} & = 0 &  & \forall i \notin A^*
\end{align}
and changing the objective function to $\max \sum_{i, j \in N} c_{ij}x_{ij} - c_{0, 2n + 1}$. Note that this last step is optional since the model already computed a route which can be paid by the served passengers, but, in practice, it is interesting to minimize the driver's cost as much as possible.

Regarding the VCG-based auctions, the original ILP can easily be modified to be used for the $\mathrm{VCG}_s$ auction as follows
    \begin{alignat*}{4}
         & \omit\rlap{$\displaystyle\max\sum_{i \in P} \sum_{j \in N} s_ix_{ij}$}                                                           \\
         & \text{subject to}                                                        & ~(\ref{constr:driver_start}) -(\ref{constr:fixed_cost}) \\
         &                                                                          & ~(\ref{constr:domain1})-(\ref{constr:domain3})
    \end{alignat*}
and also modified to be used for the $\mathrm{VCG}_r$ auction as follows%
    \begin{alignat*}{4}
         & \omit\rlap{$\displaystyle\max\sum_{i \in P} \sum_{j \in N} v_ix_{ij} - \left(\sum_{i, j \in N} c_{ij}x_{ij} - c_{0, 2n + 1}\right)$}                                                           \\
         & \text{subject to}                                                                                                                      & ~(\ref{constr:driver_start}) -(\ref{constr:fixed_cost}) \\
         &                                                                                                                                        & ~(\ref{constr:domain1})-(\ref{constr:domain3}).
    \end{alignat*}
Finally, the VCG auction can be run using the ILP for the $\mathrm{VCG}_r$ auction without Constraint~(\ref{constr:fixed_cost}).

\subsection{Brute-force algorithms}\label{brute_force}

When considering that a trip can have at most $k$ passengers and $k$ is small, one can consider brute-force algorithms for implementing the auctions presented. With this in mind, we implemented two brute force algorithms: one for VCG and $\mathrm{VCG}_r$, and another for WMS and $\mathrm{VCG}_s$.

First, notice that there are $O(n^k)$ possible trips, where $n$ is the total number of passengers, and every trip can lead to $O(2k!)$ different routes, and we are interested in the optimal one. For large $k$, a brute-force algorithm would be impractical but, in practice, one can argue that the driver cannot or does not want to pick up a large number of passengers due to vehicle space and time constraints.

The implemented brute-force algorithm for VCG and $\mathrm{VCG}_r$ enumerates all feasible trips and finds their optimal route. It, then, sorts the trips in non-increasing order by objective value. The first trip is the winner and, then, it proceeds to compute the prices. It iterates over the trips and, once the first trip~$A$ that does not contain $i$ has been found, we can compute $\max_{A \in \mathcal{A}} V_{-i}(A)$. Once this value has been computed for all $i \in A^*$, the algorithm has all the necessary information to set $p_i$ for each $i \in A^*$.

In the case of WMS and $\mathrm{VCG}_s$, the fact that the objective function does not depend on the driver's cost can be exploited to optimize the brute-force algorithm. All possible trips (including unfeasible ones) are still enumerated and sorted in non-increasing order of objective value, but feasibility and the optimal route are determined lazily when the iteration reaches a trip. This achieves better experimental performance because not all trips need to be tested for feasibility and have their cost optimized.

The algorithm is as follows. As before, the first feasible trip is the winner, and to determine $p_i$ for $i \in A^*$, the algorithm iterates over the subsequent feasible trips, skipping infeasible ones. For the $\mathrm{VCG}_s$ auction, the procedure to determine $\max_{A \in \mathcal{A}} \{ \sum_{j \in A, i \neq j} s_j\}$, is analogous to the one previously described for the VCG and~$\mathrm{VCG}_r$. For the WMS auction, both $wm^*_i$ and $A'_i$ need to be found for $i \in A^*$. The value $wm^*_i$ is obtained when the first trip that does not contain $i$ is found. To find $A'_i$, as the algorithm iterates, the maximum cardinality of a trip including~$i$ found so far is registered as \texttt{max\_cardinality[i]}. At the moment the algorithm finds a trip $B$ such that $wm(B) < wm^*_i$, then $A'_i$ is set as the current value of \texttt{max\_cardinality[i]}. The iteration terminates when $wm^*_i$ and $A'_i$ have been set for all $i \in A^*$.

\subsection{Experiments}

We have proposed three budget-balanced strategy-proof auctions for ridesharing: UMS, WMS, and $\mathrm{VCG}_s$. Also, we described the previously existing budget-balanced auction $\mathrm{VCG}_r$. As the WMS auction was superior to the UMS auction in early experiments, we will consider from now on only the WMS, the $\mathrm{VCG}_s$ and the $\mathrm{VCG}_r$ auctions.

In Section~\ref{section:setting_cost}, we proposed two ways of determining the reserve price $r_i$ of a passenger, the direct cost or the upper bound cost. We run computational experiments to gain empirical insight into the possible auction variants. We add to this comparison two non-budget-balanced auctions, the standard social welfare maximizing VCG and the Weighted Minimum Surplus (WMS) with~${r_i = 0}$~(Zero Cost) for all $i$ and no feasible trip restriction. Table~\ref{table:auctions} presents all tested auctions and the name used for them in the experiments results, and Table~\ref{table:properties} summarizes all properties that these auctions have.

\begin{table}[hbt]
    \centering
    \caption{Names used for the different auctions tested in the experiments.\label{table:auctions}}\scriptsize
    \begin{tabular}{@{}lll@{}}
        \toprule
        Auction & Reserve Price & Name         \\
        \midrule
        WMS     & Upper Bound & WMS-UB       \\
        WMS     & Direct Cost & WMS-Direct   \\
        WMS     & Zero Cost   & WMS-Zero     \\
        \midrule
        $\mathrm{VCG}_s$   & Upper Bound & $\mathrm{VCG}_s$-UB     \\
        $\mathrm{VCG}_s$   & Direct Cost & $\mathrm{VCG}_s$-Direct \\
        \midrule
        $\mathrm{VCG}_r$   & Upper Bound & $\mathrm{VCG}_r$-UB     \\
        $\mathrm{VCG}_r$   & Direct Cost & $\mathrm{VCG}_r$-Direct \\
        \midrule
        VCG     & Zero Cost   & VCG          \\
        \bottomrule
    \end{tabular}
\end{table}

\begin{table}[hbt]
    \centering
    \caption{Properties of the auctions tested. A dash represents the auction has no guarantees in the property.\label{table:properties}}\scriptsize
    \begin{tabular}{@{}lcp{2.8cm}c@{}}
        \toprule
        Name         & Budget-Balanced & Welfare                                 & Surplus Profit\\
        \midrule
        WMS-UB       & Yes             & --                                      & Lower bound\\
        WMS-Direct   & Yes             & --                                      & --  \\
        WMS-Zero     & No              & --                                      & --  \\
        \midrule
        $\mathrm{VCG}_s$-UB     & Yes             & Maximize Surplus                        & --  \\
        $\mathrm{VCG}_s$-Direct & Yes             & Maximize Surplus                        & --  \\
        \midrule
        $\mathrm{VCG}_r$-UB     & Yes             & Max among bidders with $v_i \geq r_i$ & --  \\
        $\mathrm{VCG}_r$-Direct & Yes             & Max among bidders with $v_i \geq r_i$ & --  \\
        \midrule
        VCG          & No              & Maximize                                & --\\
        \bottomrule
    \end{tabular}
\end{table}

The input instances are generated by uniformly distributing pickup and delivery locations on a rectangular section with about 100 square kilometers of area, centered on the city of São Paulo. This real-world map was obtained from the Open Street Map~\cite{OpenStreetMap} database\footnote{Map data copyrighted by OpenStreetMap contributors and available from \texttt{\url{www.openstreetmap.org}}.}. The time and distances between locations are calculated using the~OSRM~\cite{luxen-vetter-2011} library. The time constraints are that a passenger~$i$ is willing to be picked up for 15 minutes after the driver's departure time $t_0$, so~${k_i = t_0 + 15}$, and is willing to travel for at most twice the time taken in a direct route~$t_{i, i + n}$, so~$l_i = 2t_{i, i + n}$. The capacity of the vehicle is fixed at~$Q = 3$. Bids are calculated as a half-gaussian of the direct distance as in Kamar and Horvitz~\cite{kamar_collaboration_2009} such that $b_i \geq r_i$, with the standard deviation~$\sigma$ varying between instance sets. To be able to run the exact algorithms in a reasonable time, the maximum number of passengers per trip was set to 3. The ILP models were solved using the GUROBI 9 solver, and all implementations\footnote{Code and instances available at \texttt{\url{https://gitlab.com/leoyvens/ridesharing-solver2}}.} were run on a Ryzen~5~3600X CPU.\@

Tables~\ref{table:first_result}--\ref{table:last_result} show the numerical results obtained for key metrics. Each table corresponds to a combination of the total number of passengers in the instances and chosen~$\sigma$ for the bid distribution. For each combination of these parameters,~100 instances were generated and executed. There is one row per auction, and the values of each metric are the mean $\pm$ the standard deviation over all instances. Before being averaged, the values of the profit and welfare metrics were normalized by the welfare of the winning trip in the standard VCG auction. 
Table~\ref{table:runningtime} presents the running time for instances with 100 passengers and~$\sigma = 3$ when the auction is solved by ILP or by Brute Force. Even though 100 passengers could be seen as a large number of passengers competing for the trip, this number quantity serves as a stress test for the execution time of the algorithms. The run time is measured in whole seconds, therefore instances solved in under a second do not account for any run time. The best means in each metric among the budget-balanced variants are highlighted in bold.

\begin{table}[p]
    \centering
    \caption{10 passengers, $\sigma = 3$.\label{table:first_result}}
    \scriptsize
    \begin{tabular}{@{}llllll@{}}
        \toprule
        Auction                 & Profit                   & Welfare                  & Surplus Welfare          & Surplus Profit           & \# Passengers            \\
        \midrule
        VCG                     & 0.51 $\pm$ 0.28          & 1.00 $\pm$ 0.00          & 1.00 $\pm$ 0.00          & 0.51 $\pm$ 0.28          & 1.81 $\pm$ 0.46          \\
        WMS-Zero                & 0.56 $\pm$ 0.21          & 0.92 $\pm$ 0.11          & 0.92 $\pm$ 0.11          & 0.56 $\pm$ 0.21          & 1.45 $\pm$ 0.50           \\
        \midrule
        WMS-Direct              & 0.37 $\pm$ 0.26          & 0.81 $\pm$ 0.29          & 0.77 $\pm$ 0.28          & 0.33 $\pm$ 0.25          & 1.41 $\pm$ 0.64          \\
        WMS-UB                  & 0.58 $\pm$ 0.20          & 0.86 $\pm$ 0.15          & 0.71 $\pm$ 0.12          & 0.43 $\pm$ 0.19          & 1.27 $\pm$ 0.45          \\
        $\mathrm{VCG}_s$-Direct & 0.34 $\pm$ 0.29          & 0.88 $\pm$ 0.30          & 0.84 $\pm$ 0.29          & 0.30 $\pm$ 0.28          & 1.72 $\pm$ 0.67          \\
        $\mathrm{VCG}_s$-UB     & \textbf{0.59 $\pm$ 0.22} & 0.97 $\pm$ 0.08          & 0.77 $\pm$ 0.08          & 0.39 $\pm$ 0.23          & 1.67 $\pm$ 0.51          \\
        $\mathrm{VCG}_r$-Direct & 0.52 $\pm$ 0.28          & \textbf{1.00 $\pm$ 0.00} & \textbf{1.00 $\pm$ 0.00} & \textbf{0.49 $\pm$ 0.29} & \textbf{1.81 $\pm$ 0.46} \\
        $\mathrm{VCG}_r$-UB     & 0.57 $\pm$ 0.24          & \textbf{1.00 $\pm$ 0.00} & \textbf{1.00 $\pm$ 0.00} & 0.32 $\pm$ 0.24          & \textbf{1.81 $\pm$ 0.46} \\
        \bottomrule
    \end{tabular}

\end{table}
\begin{table}[p]\centering
    \caption{10 passengers, $\sigma = 5$.}
    \scriptsize
    \begin{tabular}{@{}llllll@{}}
        \toprule
        Auction                 & Profit                   & Welfare                  & Surplus Welf.            & Surplus Profit           & \# Passengers            \\
        \midrule
        VCG                     & 0.53 $\pm$ 0.28          & 1.00 $\pm$ 0.00          & 1.00 $\pm$ 0.00          & 0.53 $\pm$ 0.28          & 1.86 $\pm$ 0.47          \\
        WMS-Zero                & 0.58 $\pm$ 0.20          & 0.94 $\pm$ 0.10          & 0.94 $\pm$ 0.10          & 0.58 $\pm$ 0.20          & 1.53 $\pm$ 0.54          \\
        \midrule
        WMS-Direct              & 0.41 $\pm$ 0.27          & 0.84 $\pm$ 0.29          & 0.80 $\pm$ 0.28          & 0.37 $\pm$ 0.26          & 1.45 $\pm$ 0.64          \\
        WMS-UB                  & \textbf{0.63 $\pm$ 0.18} & 0.91 $\pm$ 0.12          & 0.75 $\pm$ 0.11          & 0.47 $\pm$ 0.17          & 1.45 $\pm$ 0.52          \\
        $\mathrm{VCG}_s$-Direct & 0.38 $\pm$ 0.29          & 0.90 $\pm$ 0.28          & 0.86 $\pm$ 0.27          & 0.34 $\pm$ 0.29          & 1.73 $\pm$ 0.66          \\
        $\mathrm{VCG}_s$-UB     & 0.60 $\pm$ 0.23          & 0.99 $\pm$ 0.03          & 0.80 $\pm$ 0.06          & 0.42 $\pm$ 0.24          & 1.76 $\pm$ 0.51          \\
        $\mathrm{VCG}_r$-Direct & 0.54 $\pm$ 0.28          & \textbf{1.00 $\pm$ 0.00} & \textbf{1.00 $\pm$ 0.00} & \textbf{0.51 $\pm$ 0.28} & \textbf{1.86 $\pm$ 0.47} \\
        $\mathrm{VCG}_r$-UB     & 0.57 $\pm$ 0.25          & \textbf{1.00 $\pm$ 0.00} & \textbf{1.00 $\pm$ 0.00} & 0.36 $\pm$ 0.24          & \textbf{1.86 $\pm$ 0.47} \\
        \bottomrule
    \end{tabular}
\end{table}
\begin{table}[p]\centering
    \caption{25 passengers, $\sigma = 3$.}
    \scriptsize
    \begin{tabular}{@{}llllll@{}}
        \toprule
        Auction                 & Profit                   & Welfare                  & Surplus Welf.            & Surplus Profit           & \# Passengers            \\
        \midrule
        VCG                     & 0.67 $\pm$ 0.19          & 1.00 $\pm$ 0.00          & 1.00 $\pm$ 0.00          & 0.67 $\pm$ 0.19          & 2.11 $\pm$ 0.31          \\
        WMS-Zero                & 0.66 $\pm$ 0.17          & 0.94 $\pm$ 0.10          & 0.94 $\pm$ 0.10          & 0.66 $\pm$ 0.17          & 1.82 $\pm$ 0.41          \\
        \midrule
        WMS-Direct              & 0.65 $\pm$ 0.19          & 0.92 $\pm$ 0.14          & 0.88 $\pm$ 0.13          & 0.61 $\pm$ 0.18          & 1.83 $\pm$ 0.40           \\
        WMS-UB                  & 0.69 $\pm$ 0.16          & 0.91 $\pm$ 0.13          & 0.74 $\pm$ 0.10          & 0.52 $\pm$ 0.14          & 1.70 $\pm$ 0.46           \\
        $\mathrm{VCG}_s$-Direct & 0.67 $\pm$ 0.21          & 0.98 $\pm$ 0.10          & 0.93 $\pm$ 0.10          & 0.62 $\pm$ 0.21          & 2.08 $\pm$ 0.34          \\
        $\mathrm{VCG}_s$-UB     & \textbf{0.72 $\pm$ 0.17} & 0.99 $\pm$ 0.03          & 0.79 $\pm$ 0.04          & 0.52 $\pm$ 0.17          & 2.00 $\pm$ 0.35           \\
        $\mathrm{VCG}_r$-Direct & 0.67 $\pm$ 0.19          & \textbf{1.00 $\pm$ 0.00} & \textbf{1.00 $\pm$ 0.00} & \textbf{0.62 $\pm$ 0.19} & \textbf{2.11 $\pm$ 0.31} \\
        $\mathrm{VCG}_r$-UB     & 0.68 $\pm$ 0.19          & \textbf{1.00 $\pm$ 0.00} & \textbf{1.00 $\pm$ 0.00} & 0.45 $\pm$ 0.19          & \textbf{2.11 $\pm$ 0.31} \\
        \bottomrule
    \end{tabular}
\end{table}
\begin{table}[p]\centering
    \caption{25 passengers, $\sigma = 5$.}
    \scriptsize
    \begin{tabular}{@{}llllll@{}}
        \toprule
        Auction                 & Profit                   & Welfare                  & Surplus Welfare          & Surplus Profit           & \# Passengers            \\
        \midrule
        VCG                     & 0.69 $\pm$ 0.17          & 1.00 $\pm$ 0.00          & 1.00 $\pm$ 0.00          & 0.69 $\pm$ 0.17          & 2.09 $\pm$ 0.32          \\
        WMS-Zero                & 0.68 $\pm$ 0.15          & 0.96 $\pm$ 0.09          & 0.96 $\pm$ 0.09          & 0.68 $\pm$ 0.15          & 1.88 $\pm$ 0.36          \\
        \midrule
        WMS-Direct              & 0.68 $\pm$ 0.19          & 0.94 $\pm$ 0.14          & 0.90 $\pm$ 0.13          & 0.63 $\pm$ 0.18          & 1.87 $\pm$ 0.37          \\
        WMS-UB                  & \textbf{0.73 $\pm$ 0.12} & 0.95 $\pm$ 0.10          & 0.78 $\pm$ 0.08          & 0.56 $\pm$ 0.11          & 1.83 $\pm$ 0.40          \\
        $\mathrm{VCG}_s$-Direct & 0.68 $\pm$ 0.19          & 0.99 $\pm$ 0.10          & 0.94 $\pm$ 0.10          & 0.64 $\pm$ 0.18          & 2.05 $\pm$ 0.33          \\
        $\mathrm{VCG}_s$-UB     & 0.73 $\pm$ 0.14          & 0.99 $\pm$ 0.03          & 0.81 $\pm$ 0.03          & 0.54 $\pm$ 0.14          & 2.03 $\pm$ 0.33          \\
        $\mathrm{VCG}_r$-Direct & 0.69 $\pm$ 0.16          & \textbf{1.00 $\pm$ 0.00} & \textbf{1.00 $\pm$ 0.00} & \textbf{0.64 $\pm$ 0.17} & \textbf{2.09 $\pm$ 0.32} \\
        $\mathrm{VCG}_r$-UB     & 0.69 $\pm$ 0.16          & \textbf{1.00 $\pm$ 0.00} & \textbf{1.00 $\pm$ 0.00} & 0.50 $\pm$ 0.17          & \textbf{2.09 $\pm$ 0.32} \\
        \bottomrule
    \end{tabular}
\end{table}

\begin{table}[p]
    \centering
    \caption{50 passengers, $\sigma = 3$.}

    \scriptsize
    \begin{tabular}{@{}llllll@{}}
        \toprule
        Auction                 & Profit                  & Welfare                 & Surplus Welfare          & Surplus Profit           & \# Passengers            \\
        \midrule
        VCG                     & 0.78 $\pm$ 0.15         & 1.00 $\pm$ 0.00         & 1.00 $\pm$ 0.00          & 0.78 $\pm$ 0.15          & 2.29 $\pm$ 0.46          \\
        WMS-Zero                & 0.75 $\pm$ 0.10         & 0.95 $\pm$ 0.07         & 0.95 $\pm$ 0.07          & 0.75 $\pm$ 0.10          & 1.98 $\pm$ 0.20           \\
        \midrule
        WMS-Direct              & 0.76 $\pm$ 0.11         & 0.95 $\pm$ 0.07         & 0.90 $\pm$ 0.07          & 0.72 $\pm$ 0.11          & 2.00 $\pm$ 0.20            \\
        WMS-UB                  & 0.78 $\pm$ 0.11         & 0.95 $\pm$ 0.08         & 0.77 $\pm$ 0.07          & 0.60 $\pm$ 0.10          & 1.94 $\pm$ 0.28          \\
        $\mathrm{VCG}_s$-Direct & 0.79 $\pm$ 0.13         & 1.00 $\pm$ 0.01         & 0.95 $\pm$ 0.02          & \textbf{0.74 $\pm$ 0.14} & \textbf{2.29 $\pm$ 0.46} \\
        $\mathrm{VCG}_s$-UB     & \textbf{0.80 $\pm$ 0.11}& 1.00 $\pm$ 0.01         & 0.81 $\pm$ 0.03          & 0.61 $\pm$ 0.12          & 2.23 $\pm$ 0.42          \\
        $\mathrm{VCG}_r$-Direct & 0.78 $\pm$ 0.14         & \textbf{1.00 $\pm$ 0.00}& \textbf{1.00 $\pm$ 0.00} & 0.73 $\pm$ 0.15          & \textbf{2.29 $\pm$ 0.46} \\
        $\mathrm{VCG}_r$-UB     & 0.79 $\pm$ 0.14         & \textbf{1.00 $\pm$ 0.00}& \textbf{1.00 $\pm$ 0.00} & 0.59 $\pm$ 0.14          & \textbf{2.29 $\pm$ 0.46} \\
        \bottomrule
    \end{tabular}
\end{table}

\begin{table}[p]\centering
    \caption{100 passengers, $\sigma = 3$.\label{table:last_result}}
    \scriptsize
    \begin{tabular}{@{}llllll@{}}
        \toprule
        Auction                 & Profit                   & Welfare                 & Surplus Welfare          & Surplus Profit           & \# Passengers           \\
        \midrule
        VCG                     & 0.82 $\pm$ 0.12          & 1.00 $\pm$ 0.00         & 1.00 $\pm$ 0.00          & 0.82 $\pm$ 0.12          & 2.52 $\pm$ 0.50          \\
        WMS-Zero                & 0.79 $\pm$ 0.08          & 0.95 $\pm$ 0.05         & 0.95 $\pm$ 0.05          & 0.79 $\pm$ 0.08          & 2.00 $\pm$ 0.14          \\
        \midrule
        WMS-Direct              & 0.80 $\pm$ 0.08          & 0.95 $\pm$ 0.05         & 0.90 $\pm$ 0.05          & 0.75 $\pm$ 0.08          & 2.00 $\pm$ 0.14          \\
        WMS-UB                  & 0.81 $\pm$ 0.08          & 0.95 $\pm$ 0.06         & 0.78 $\pm$ 0.05          & 0.64 $\pm$ 0.07          & 2.00 $\pm$ 0.14          \\
        $\mathrm{VCG}_s$-Direct & 0.82 $\pm$ 0.11          & \textbf{1.00 $\pm$ 0.00}& 0.95 $\pm$ 0.01          & \textbf{0.77 $\pm$ 0.11} & \textbf{2.52 $\pm$ 0.50} \\
        $\mathrm{VCG}_s$-UB     & \textbf{0.84 $\pm$ 0.09} & 1.00 $\pm$ 0.01         & 0.81 $\pm$ 0.02          & 0.65 $\pm$ 0.10          & 2.44 $\pm$ 0.50          \\
        $\mathrm{VCG}_r$-Direct            & 0.82 $\pm$ 0.12          & \textbf{1.00 $\pm$ 0.00}& \textbf{1.00 $\pm$ 0.00} & 0.76 $\pm$ 0.12          & \textbf{2.52 $\pm$ 0.50} \\
        $\mathrm{VCG}_r$-UB                & 0.82 $\pm$ 0.12          & \textbf{1.00 $\pm$ 0.00}& \textbf{1.00 $\pm$ 0.00} & 0.63 $\pm$ 0.12          & \textbf{2.52 $\pm$ 0.50} \\
        \bottomrule
    \end{tabular}
\end{table}

\begin{table}[p]\centering
    \caption{Algorithms' running time for 100 passengers and $\sigma = 3$.\label{table:runningtime}}
    \scriptsize
    \begin{tabular}{@{}lll@{}}
        \toprule
        Auction                 & ILP (s)                 & Brute Force (s)           \\
        \midrule
        VCG                     & 923.41 $\pm$ 827.46          & 0.45 $\pm$ 0.22          \\
        WMS-Zero                & 3959.5 $\pm$ 6353.58         & 0.07 $\pm$ 0.05          \\
        \midrule
        WMS-Direct              & 3117.53 $\pm$ 5898.95        & 0.08 $\pm$ 0.05          \\
        WMS-UB                  & 1276.05 $\pm$ 2517.95        & \textbf{0.07 $\pm$ 0.04} \\
        $\mathrm{VCG}_s$-Direct & 1186.15 $\pm$ 1144.35        & 0.45 $\pm$ 0.22          \\
        $\mathrm{VCG}_s$-UB     & \textbf{493.52 $\pm$ 464.65} & 0.46 $\pm$ 0.23          \\
        $\mathrm{VCG}_r$-Direct            & 960.66 $\pm$ 855.65          & 0.46 $\pm$ 0.22          \\
        $\mathrm{VCG}_r$-UB                & 1053.1 $\pm$ 932.22          & 0.45 $\pm$ 0.21          \\
        \bottomrule
    \end{tabular}
\end{table}

Regarding welfare, the $\mathrm{VCG}_r$ auction was the clear winner, always obtaining the same welfare as the (non-budget-balanced) VCG auction.
Recall that $\mathrm{VCG}_r$ eliminates bidders with low bids ($v_i < r_i$) and considers only alternatives $A$ such that ${cost(A) \leq \sum_{i \in A} r_i}$. Thus, in our experiments it is not unexpected that VCG and $\mathrm{VCG}_r$ have the same welfare as VCG will not select bidders with low bids, which themselves have a low direct cost and, in turn, a low $r_i$, as both reserve price functions are dependent on the direct cost. 
Clearly, this not always true, since you can have bidders with high reserve prices and low bids, that would discard by $\mathrm{VCG}_r$ but would still be selected by VCG\@.

The $\mathrm{VCG}_s$ auction also performed well, since it always obtained mean welfare over 88\% of the maximum possible welfare. On the instance sets of 50 and 100 passengers, the $\mathrm{VCG}_s$-Direct and $\mathrm{VCG}_s$-UB variants even obtained a mean that is indistinguishable from the theoretical maximum. The results regarding the $\mathrm{VCG}_r$ and the $\mathrm{VCG}_s$ auctions are impressive, considering that the budget-balanced variants have no guarantee of maximizing social welfare. Finally, the WMS auctions obtained lower welfare than the VCG ones, which is expected since the WMS auction does not directly optimize for welfare, but still it had mean welfare over 90\% of the maximum on all instance sets larger than 10 passengers. Therefore, we can see that all budget-balanced auctions obtained a high proportion of the maximum welfare, even without the theoretical guarantee of maximizing social welfare.

For the passengers metric, which is the number of served passengers, the~VCG variants consistently obtained a higher mean than the WMS ones. This is because adding more passengers usually increases welfare. Consequently, improving the VCG objective function. However, in WMS auctions, adding a passenger with a low bid may reduce the value of the WMS auction objective~function.

Comparing the profit of VCG variants and WMS variants, we can see that~WMS variants did slightly better on the 10 passenger instances, while the VCG variants did slightly better on the larger instance sets. But, without exception, when comparing for the same number of passengers, the same $\sigma$ and the same reserve price variant, the mean of the VCG variants was always within half a standard deviation of the WMS variant, and vice-versa. Therefore, we can claim that VCG and WMS variants had equivalent results of profit in this~experiment.

Observing the run time,
Table~\ref{table:runningtime} shows the high computational cost of running the auction using an ILP solver when we have 100 passengers. Nonetheless, brute-force algorithms were able to produce solutions in less than one second on average. If we increase the number of passengers that the driver can pick up, the brute-force algorithm will start to perform poorly. Nonetheless, we do not expect that, in practice, many passengers will be served by a single driver either due to car capacity, time constraints, or the inconvenience of doing so. Finally, notice that, due to optimizations previously described, the brute-force algorithm for the WMS auction was able to run much faster than the VCG-based~counterparts.

The instance sets of 10 and 25 passengers were run with $\sigma = 3$ and ${\sigma = 5}$, so we can measure the effect of different standard deviation values for the half-gaussian distribution of bids. We see that for both instances sizes for all auctions, the instance set with $\sigma = 5$ achieved a better or equal value of profit and welfare, so all auctions were able to take advantage of the higher bids.

We can also observe the results of the different reserve price functions. It is important to note that the non-budget-balanced auctions VCG and WMS-Zero did incur a loss in some instances, so the budget-balanced property has practical consequences. Looking at the profit, it is interesting to notice that in the~10~passenger instances the direct cost except for $\mathrm{VCG}_r$ performed worse than the zero cost. In instances with more passengers, these two reserve price types had similar performance. In these experiments, the upper bound cost always had a higher mean profit than the other two reserve price functions, by setting a higher fixed reserve price it was able to charge higher prices from the passengers with higher bids. We also noticed that in general, a higher reserve price function had as a consequence a lower mean runtime, with the zero cost being the slowest and the upper bound the fastest, this can be explained by higher reserve prices filtering out more passengers whose bid is lower than their reserve price.

Overall, the parameters that most significantly increase the empirical profit and welfare are a higher number of passengers and a bid distribution with a higher $\sigma$. Note that it is not simply stating that more passengers and higher bids mean higher raw numbers. We analyze normalized values, so all these auctions are in fact more efficient in obtaining a higher ratio of the maximum achievable profit and welfare when there are more passengers in the instance. This is in line with other experimental results in the literature such as Kamar and Horvitz~\cite{kamar_collaboration_2009} and Kleiner et al.~\cite{kleiner_mechanism_2011}, whose systems were also more efficient with a higher number of passengers. The choice of reserve price function also had a significant effect, so the reserve price function should be tailored to the situation in which the auction is being applied. 

\section{Multiple Drivers}\label{sec:multidriver}

In practice, a ride-sharing system will be used in a multi-driver scenario. Thus, in this section, we consider possible adaptations of the proposed auctions to this environment.

We can adapt $\mathrm{VCG}_r$ and $\mathrm{VCG}_s$ to optimize the system, considering all drivers at once. To do so, one must select one winning trip $A_d^*$ for every driver~$d$ maximizing the social welfare or the surplus welfare, respectively. It is also necessary that, for every driver $d$, $\sum_{i \in A^*_d} r_i \geq \mathrm{\emph{cost}}_d(A^*_d)$ where $\mathrm{\emph{cost}}_d(A)$ is the cost of driver $d$ to serve $A$. The mechanism will be budget-balanced and also strategy-proof. Unfortunately, defining $r_i$ becomes harder as there are multiple drivers. For example, we can use the direct cost but not the upper bound cost, as it is driver-dependent.

We can also adapt the WMS auction to the multiple-driver scenario, considering that an alternative is the set of served passengers among all drivers. In this case, alternative $A$ would be feasible if it can be partitioned in one trip~$A_d^*$ for each driver $d$ such that $\sum_{i \in A^*_d} r_i \geq \mathrm{\emph{cost}}_d(A^*_d)$. Again, the mechanism would be budget-balanced and strategy-proof. All definitions presented for the WMS auction to select the winning alternative and to define the prices to be paid would be the same. 

One issue of considering these auctions for the multi-driver scenario is that, as the number of drivers increases, so do the computational effort necessary for finding an optimal solution, which is mandatory for these auctions. In this case, one could either solve the problem heuristically and lose strategy-proofness, design fast exact algorithms using the several techniques known for routing problems (which would, probably, still be very challenging) or develop other auctions that do not need that an optimal solution is found (such as auctions that use a greedy heuristic~\cite{Lehmann02}), again using the idea of reserve prices to guarantee that the mechanism is budget-balanced.
The problem could also be solved in a parallelized way if we predefine (independently of the bid) which driver could serve which passengers using a clustering algorithm\footnote{Considering that the passengers cannot lie their origin and destination, we would still have a strategy-proof auction in this case.}.

Another issue is that, in practice, passengers and drivers arrive at the system in an online fashion and the proposed mechanism only looks at a snapshot of the situation.

Regarding this difficulty, one possibility is to run the mechanism from time to time. But, in this case, the passenger could give a smaller bid to not be served in a more competitive round and then be served in a following less competitive round, which could make them pay less to be served. The passenger could also wait a little while to enter the system in another run that would reduce the price paid. That is, the auction loses its strategy-proofness when we look at it as a repeated auction (but it is still budget-balanced). Thus, in the multi-driver scenario, an interesting research subject would be to develop strategy-proof budget-balanced online auctions. Notice that this is not a problem for the single-driver scenario, as the passengers that arrive after the driver leaves his starting location do not partake in the auction and the auctions are sealed-bid.

If we use any of the proposed auctions in this paper greedily, choosing which passengers to serve one driver at a time, we would not have strategy-proofness, but we could argue that the auction would still be ``myopically'' strategy-proof. That is, if we consider that the passengers cannot take into account other drivers that will appear later on because they are unaware that they will be available, it would still be in the passenger's best interest to bid truthfully. Fortunately, as the experiments show, each single auction can be solved fast using a brute-force algorithm even for a large number of passengers.

\section{Conclusion and future research}\label{sec:conclusion}

The main idea of this paper is that, through the usage of passengers reserve prices, we can obtain strategy-proofness and budget balancedness as long as the reserve prices do not depend on the passengers bids. Using this concept, we analyze the theoretical and empirical properties of two newly proposed budget-balanced strategy-proof auctions for ridesharing problems, the $\mathrm{VCG}_s$ auction and the WMS auction.

Even if the proposed auctions are better suited for the single-driver scenario, they can be used in a greedy fashion while maintaining budget balancedness and still provide some degree of strategy-proofness (if we consider that the users are unaware of the other drivers). Also, the idea of using passengers reserve prices can be a stepping stone for mechanisms for the multi-driver scenario, in particular, when the drivers and passengers arrive in an online-fashion.

The experimental results, which utilized randomly generated instances and also includes a VCG auction with bidder reserve prices (named $\mathrm{VCG}_r$) for comparison, showed equivalence between the WMS auction and the VCG auctions in the welfare and profit metrics. 

However, the WMS auction running time, which is an important factor in practice, was around 18\% of the running time of the other methods. It also
offers interesting theoretical results with the lower bounds for the surplus welfare and surplus profit which we proved in Theorems~\ref{theo:approx},~\ref{theo:profit} and~\ref{theo:ratio}, under the downward closed alternatives' assumption.
This is an important feature, as VCG-based mechanisms can have surplus profit close or even equal to zero in low-competition scenarios.

The WMS auction can be applied not only to ridesharing, but to any single-parameter domain in which individual bidders form groups, analogous to the trips in ridesharing, and there are complex restrictions on what groups are allowed to be formed. Applying the WMS auction to problems outside ridesharing is a possibility for future research, particularly problems with downward closed~alternatives. 

\section*{Acknowledgments}
This study was supported by Grants
{425340/2016--3}, 
{308689/2017--8}, 
\mbox{425806/2018--9}, 
and
{311039/2020--0} 
from the National Council for Scientific and Technological Development (CNPq), and Grant 
\mbox{2015/11937--9} 
from the São Paulo Research Foundation (FAPESP).

\bibliographystyle{elsarticle-num}
\bibliography{artigo-clean}

\clearpage
\appendix

\section{Weighted vs. Unweighted Minimum Surplus}\label{appendix:ums}
The following result shows that in many cases the UMS auction cannot achieve a better \emph{surplus payment} than the WMS auction. The surplus payment~$sp(A^*)$ is the amount paid by the passengers discounting the reserve price so ${sp(A) = \sum_{i \in A} (p_i - r_i)}$. This surplus payment is meaningful as a lower bound on the profit which is ${\mathit{profit}(A) = \sum_{i \in A} p_i - \mathrm{\emph{cost}}(A)}$ since~${\sum_{i \in A} r_i \geq \mathrm{\emph{cost}}(A)}$ in a feasible trip.

\begin{theorem}
    For a given set of trips $\mathcal{A}$ let $A^*$ be the winner in the weighted variant and let~$A'^*$ be the winner in the unweighted variant. If $A^* \cap A'^* = \emptyset$ and $|A'_i| = |A^*|$ for all $i \in A^*$, then~${sp(A^*) \geq sp(A'^*)}$.
\end{theorem}

\begin{prf}
    By hypothesis, if $i \in A^*$ then $i \not\in A'^*$ so~$wm^*_{i} \geq |A'^*|s_{\min}(A'^*)$. From these two facts we get
    \[
        sp(A^*) \geq~\sum_{i \in A^*}  \frac{|A'^*|s_{\min}(A'^*)}{|A'_i|}
    \]
    Since we assume $|A'_i| = |A^*|$ we may simplify to~$sp(A^*) \geq |A'^*|s_{\min}(A'^*)$ by observing that the dividend~$|A'^*|s_{\min}(A'^*)$ does not depend on $i$, and that $\sum_{i \in A^*} \frac{1}{|A^*|} = 1$. In the unweighted variant the surplus payment is~$sp(A'^*) = \sum_{i \in A^*} ss'_i$, now since~${s_{\min}(A'^*) \geq ss'_i}$ for all $i \in A'^*$ we have~${|A'^*|s_{\min}(A'^*) \geq sp(A'^*)}$. Joining the two inequalities, we conclude that~${sp(A^*) \geq sp(A'^*)}$.\qed
\end{prf}

\section{Results for the WMS auction}\label{appendix:wms}
This section proves some technical results regarding the WMS auction.

\begin{lemma}
    If there is $v_i \in \mathbb{R}$ such that $i\in A^*$ for some $v_i$, then $A'_i$ exists.\label{cant_win}
\end{lemma}
\begin{prf}
    We prove the contrapositive: If there is no $A'_i$ then $i \not\in A^*$ for any $v_i$. By the definition we have that $A'_i$ does not exist if and only if $\{ i \} \not\in \mathcal{A}$ and for all $A \in \mathcal{A}$
    \begin{equation}
        \label{A_i_undef}
        i \in A \implies wm^*_{i} > |A| s_{\min}(A \setminus \{i \}).
    \end{equation}
    We must show that this implies $i \not\in A^*$. Considering more elements can only lower the minimum, so from~(\ref{A_i_undef}) we have $wm^*_{i} > wm(A)$ for all $A$ with $i \in A$. Since $A^*$ is the winner we know that~${wm(A^*) \geq wm^*_{i}}$, joining the inequalities gives $wm(A^*) > wm(A)$ for any $A$ such that $i \in A$, and we conclude that $i \not\in A^*$.\qed
\end{prf}

\begin{theorem}
    The bound on the surplus welfare ratio of $H_{\max |A|}$ given in Theorem~\ref{theo:approx} is tight.
\end{theorem}

\begin{prf}
    To prove that the bound is tight, we construct $\mathcal{A}$ such that the surplus welfare ratio is~$H_{\max |A|}$. Let~$B$ be a feasible alternative with an arbitrary~$|B|$ and~$wm(B)$, and let~$s_i = \frac{wm(B)}{i}$ for the~$i$-th passenger in~$B$. Let $C$ be a feasible alternative such that~${|C| = |B|}$ and~${s_i = \frac{wm(B)}{|C|}}$ for all~${i \in C}$. Assume that in a tie-break~$C$ is preferred over any other alternative. Let ${\mathcal{A} = P(B) \cup P(C)}$ where~$P(A)$ is the power set of $A$, that is, the set of all subsets, so $\mathcal{A}$ is downward~closed.

    We now show that $A^* = C$. $A^* \in \mathcal{A}$, so $A^* \in P(B) \cup P(C)$. All passengers in~$C$ have the same~$s_i$, so all~$C' \in P(C)$ have the same $s_{\min}(C')$, but $C$ has the largest cardinality, so for~$C' \in P(C)$ the best alternative is~$C$. This narrows the possible winners to $A^* \in P(B) \cup \{ C \}$. For $B' \subseteq B$, we have that~${s_{\min}(B') \leq \frac{wm(B)}{|B'|}}$ so $wm(B') \leq wm(B)$. But $wm(C) = wm(B)$ and $C$ wins any tie-break, therefore $A^* = C$.

    If $A' \in P(A)$ we have that $V_s(A') \leq V_s(A)$ so $\max_{A \in \mathcal{A}} \{ V_s(A) \}$ must be~$V_s(B)$ or~$V_s(C)$. Noting that~${V_s(B) = wm(B)H_{|B|}}$ and that $V_s(C) = wm(B)$, we have~${\max_{A \in \mathcal{A}} \{ V_s(A) \}= V_s(B)}$. Combining the equations $\max_{A \in \mathcal{A}} \{ V_{s}(A) \} = V_s(C)H_{|B|}$, $\max_{A \in \mathcal{A}} \{ |A| \} = |B|$, and $A^* = C$, we see that
    \[
        \frac{\max_{A \in \mathcal{A}} \{ V_{s}(A) \}}{V_{s}(A^*)} = H_{\max |A|}.
    \]
    Therefore, the bound is tight.\qed
\end{prf}

\end{document}